\documentclass[a4paper,11pt]{article}

\usepackage{jcappub} 

\usepackage[T1]{fontenc} 

\title{\boldmath A thermodynamic point of view on dark energy models}

\author[a,1]{V.F. Cardone\note{Corresponding author},}
\author[b,c]{N. Radicella,}
\author[b]{A. Troisi}

\affiliation[a]{I.N.A.F.\,-\,Osservatorio Astronomico di Roma, via Frascati 33, 00040 - Monte Porzio Catone (Roma), Italy}
\affiliation[b]{Dipartimento di Fisica ``E.R. Caianiello", Universit\`{a} di Salerno, Via Giovanni Paolo II, 132, 84084 - Fisciano (Sa), Italy}
\affiliation[c]{I.N.F.N. - Sez. di Napoli - GC di Salerno, Via Giovanni Paolo II, 132, 84084 - Fisciano (Sa), Italy}

\emailAdd{winnyenodrac@gmail.com}
\emailAdd{ninfa.radicella@gmail.com}
\emailAdd{antrois@gmail.com}

\abstract{
We present a conjugate analysis of two different dark energy models investigating both their agreement with recent data and their thermodynamical properties. The successful match with the data allows to both constrain the model parameters and characterize their kinematical properties. As a  novel step, we exploit the strong connection between gravity and thermodynamics to further check models viability by investigating their thermodynamical quantities. In particular, we study whether the cosmological scenario fulfills the generalized second law of thermodynamics and, moreover, we contrast the two model asking whether the evolution of the total entropy is in agreement with the expectation for a closed system.
As a general result, we discuss whether thermodynamic constraints can be a valid complementary way to both constrain dark energy models and differentiate among rival scenarios.}

\begin{document}
\maketitle
\flushbottom

\section{Introduction}

That the universe is presently undergoing a phase of accelerated expansion is nowadays an observationally unquestionable fact \cite{RiessPerlmutter,SNeIa-2,SNeIa-3,SNeIa-4, SNeIa-5}. Understanding what is driving this speed up is one of most debated issue of modern cosmology with many contenders on the ground. An apparently simple solution to this fascinating problem may be found in the framework of Einsteinian General Relativity relying on the cosmological constant. Indeed, the $\Lambda$CDM model excellently reproduces the most recent observational data \cite{planck} assigning to the $\Lambda$ term the origin of cosmic acceleration, while cold dark matter (CDM) is in charge for assembling the large scale structure. Unfortunately, the remarkable success in fitting the data comes at a non negligible price: the inferred $\Lambda$ being 120 orders of magnitude larger than the predicted one with no idea of what cancellation mechanism can cure such a spectacular discordance. As a further issue, it is also disturbing that $\Lambda$ and CDM contribute roughly equally to the today total energy budget which is quite unlikely considering that this only occurs once during the universe evolutionary history. Replacing $\Lambda$ with an exotic fluid with a varying negative pressure, commonly referred to as dark energy (DE), still does not solve the coincidence problem  (see, e.g., \cite{sahnipeebles,burgess2013,deputter2010} for this and other related issues). 

Lacking a solid theoretical basis, in the last decade, it has become popular to rely on phenomenological models~\cite{DEreview}. On one hand, DE is modelled as a perfect barotropic fluid with pressure $p$ and energy density $\rho$ related by $p = w \rho$ with $w$ the equation of state which becomes the uncertain feature assigning the model. Noting that $w = -1$ identifies $\Lambda$CDM model, the immediate next step is a Taylor expansion in the scale factor $a$ thus leading to $w = w_0 + w_a (1 - a)$ which is known as the Chevallier\,-\,Polanski\,-\,Linder parameterization \cite{chevallier01,linder03}. Such a simple formula is nevertheless able to mimic reasonably well the expansion rate of a large class of theoretically motivated DE models making it the reference choice for dark energy. Indeed, it is now common to quantify the capability of an observable to constrain DE in terms of the expected errors on the $(w_0, w_a)$ parameters. On the other hand, one can note that, in a Friedman\,-\,Robertson\,-\,Walker (FRW) framework, the expansion rate is directly determined by the energy density of the fluids composing the universe. As such, one can therefore directly model $\rho$ instead of $w$ with the further attracting possibility to work out models which can mimic both dark energy and dark matter with a single term. This is the case of the {\it Hobbit} model proposed by some of us \cite{cardone04} with an energy density scaling with $a$ in such a way that the fluid evolves first as radiation, then as CDM and finally as a cosmological constant. 

Despite the dynamical dark energy models capability to interpret cosmic expansion and the wide versatility of such models with respect to other experimental prescriptions, e.g. the perturbations ``realm" \cite{dynPert2015}, one should also be aware that the proposed scenarios are not in conflict with the prescriptions descending from fundamental physics. To this end, one can exploit the intriguing connection between gravity and thermodynamics which is well illustrated by the case of black holes (hereafter, BHs). In the semiclassical description of their physics, BHs behave as black bodies emitting thermal radiation and have both a temperature \cite{hawking} and an entropy \cite{bekenstein}. According to the Hawking\,-\,Bekenstein formula, the event horizon entropy is proportional to its area through the Planck and Boltzmann constants. Moreover, it has also been shown that BHs in contact with their own radiation satisfy the second law of thermodynamics. These results have been generalized in a cosmological context considering the universe as an isolated thermodynamic system filled with a cosmic fluid and bounded by some horizon \cite{wang06,izquierdo2006}. Although this connection has been later reinforced \cite{jacobson1995,padmanabhan2005}, the deep relation between gravity, thermodynamics and quantum mechanics is far from being fully unveiled. Nevertheless, one can pursue this intriguing link as long as the universe can be considered a closed system with a horizon. In analogy with classical thermodynamics, one should expect that it naturally evolves towards a thermodynamical equilibrium. Therefore, its overall entropy function $S(x)$, being $x$ the relevant variable, should be a positively defined quantity that increases with convex concavity \cite{callen1960}. In practice, any physically viable cosmological model needs to fulfill the two conditions $dS(x)/dx\geq 0$  and, in the far future, $d^2S(x)/dx^2 < 0$. In particular, one can consider to evaluate the Universe effective entropy amount (including both the horizon contribution and the matter\,-\,energy term) and check that its first and second derivative have the correct behaviour. It could then be possible to discriminate among thermodynamically suitable cosmological models by considering the effective contribution of the different components to the entropy evolution \cite{radicella10,radicella11,radicella12,radicella13,radicella14}.

In the standard approach to dark energy models, one is typically only concerned with its kinematic and dynamical properties so that the viability of a given parameterization is judged on the basis of how well it reproduces the observed data. We advocate here the need for a next step on. We consider two different dark energy models and constrain their parameters fitting a wide range of up to date observations probing the expansion rate. We then move on asking whether the models also comply with the second law of thermodynamics. Put in other words, we do not halt at the observations level, but push the analysis a step further into the realm of thermodynamics. As we will see, this allows to put stronger constraints on the model parameters and discriminate them at a more fundamental level.

Plan of the paper is as follows. In Section II, we resume the basics of the two cosmological models we consider, while Section III is devoted to matching them with observations. Section IV presents a general discussion on thermodynamical relations of both the horizon and the fluids making the cosmic pie, while Section V demonstrates the potentiality of a thermodynamical analysis applying it to the two models previously matched with the data. A summary of the results and further prospectives are finally given in Section VI.

\section{Dark energy models}\label{sec: model}

In the framework of General Relativity and a homogenous and isotropic spatially flat universe described by the Robertson\,-\,Walker metric, the cosmic background evolution is determined by the Friedman equations, the first one being\,:

\begin{equation}
H^2(a) = \frac{8 \pi G}{3} \sum{\rho_i(a) }  \ ,
\label{friedmann}
\end{equation}
where $H = \dot{a}/a$ is the Hubble expansion rate, $a$ the scale factor (with present day value $a_0 = 1$), the dot denotes derivative with respect to cosmic time, and the sum is over the fluids contributing to the total energy budget. We will hereafter use the labels $(R, M, DE)$ to refer to the radiation, matter, and dark energy terms. The second Friedmann equation may be replaced by the continuity equation which sets the scale factor dependence of the density. Assuming no interaction between fluids, we can conveniently write

\begin{equation}
\frac{d\ln{\rho_i(a)}}{d\ln{a}} = -3 [(1 + w_i(a)]
\label{eq: conteq}
\end{equation}
where $w_i(a)$ is the equation of state (EoS) of the $i$\,-\,th term. Setting $w_R = 1/3$ for radiation and $w_M = 0$ for dust matter, we get

\begin{equation}
E^2(a) = \frac{H^2(a)}{H_0^2} = \Omega_M a^{-3} + \Omega_R a^{-4} 
+\Omega_{DE} f_{DE}(a)
\label{eq: evsa}
\end{equation}
with 
\begin{equation}
\Omega_i = \frac{\rho_i(a = 1)}{\rho_{crit}(a = 1)} = \left ( \frac{3 H_0^2}{8 \pi G} \right )^{-1} \rho_i(a = 1) \ ,
\end{equation}
and $\Omega_{DE} = 1 - \Omega_R - \Omega_M$ because of the flatness condition.  The density parameters\footnote{Hereafter, we denote with $\Omega_i$ the present day value of the density parameter for the $i$\,-\,th component, while $\Omega_i(a)$ is its scale factor dependent version.} will then read

\begin{displaymath}
\Omega_M(a) = \frac{\Omega_M a^{-3}}{E^2(a)} \ \ , \ \
\Omega_R(a) = \frac{\Omega_R a^{-4}}{E^2(a)} \ \ , \ \ 
\Omega_{DE}(a) = \frac{\Omega_{DE} f_{DE}(a)}{E^2(a)} \ \ , 
\end{displaymath}
with $f_{DE}(a) = \rho_{DE}(a)/\rho_{DE}(a = 1)$ as obtained by integrating Eq.(\ref{eq: conteq}) given an expression for the DE EoS.  

For later convenience, it is worth noting that the second Friedmann and the continuity equations can be suitably rewritten as follows

\begin{eqnarray}
\frac{H'}{H}&=&-\frac{3}{2a}\left[1+\sum_i w_i\Omega_i (a)\right]\label{secondFr}\\
\Omega'_i(a)&=&\frac{3\  \Omega_i(a)}{a}\left[w_i+\sum_j w_j\Omega_j(a)\right].\label{continuity}\ ,
\end{eqnarray}
where prime denotes derivative with respect to $a$.

\subsection{The BA parameterization}

The ignorance about the nature of dark energy has motivated the search for analytical expressions which can mimic the DE EoS of a vast class of theoretical models. The most popular one is by far the one referred to as the CPL parametrization \cite{chevallier01,linder03}:
\begin{equation}
w_{DE}(a) = w_{CPL}(a) = w_0 + w_a (1-a)
\label{eq: wcpl}
\end{equation}
where $w_0$ is the present day value and $w_a$ sets the first derivative. Note that, while the early universe limit is well defined with $w_{DE}(a \rightarrow 0) = w_0 + w_a$, the EoS diverges asymptotically in the future $(a \rightarrow \infty)$. Notwithstanding this problem, the CPL model is routinely used in the literature. However, we will be interested also in the future time evolution of the DE EoS so that we need a model which is as similar as possible to the CPL one, but avoid any singularity both in the early universe and the asymptotic future. This can be accomplished resorting to the parameterization proposed by Barbosa \& Alcaniz \cite{ba} setting

\begin{equation}
w_{DE}(a) = w_{BA}(a) = w_0 + w_a (1 - a) [1 - 2 a (1 - a)]^{-1} \ ,
\label{eq: wba}
\end{equation}
which we will refer to in the following as the BA model. It is immediate to show that the BA EoS asymptotes to $w_0 + w_a$ for $a \rightarrow 0$ as for the CPL case, but it is not divergent in the future where the EoS goes to $w_0$. Integrating Eq.(\ref{eq: conteq}), we then get\,:

\begin{equation}
f_{DE}(a) =  f_{BA}(a) = a^{-3 (1 + w_0)} \left [ 1 + \left (\frac{1 - a}{a} \right )^2 \right ]^{-3 w_a/2} \ ,
\label{eq: rhoba}
\end{equation}
which sets the evolution of the DE for this model.

\subsection{The Hobbit model}

Rather than assuming an expression for the DE EoS, one can also use Eq.(\ref{eq: conteq}) in the reverse way, i.e. as a straightforward derivation of $w_{DE}(a)$ given an analytical expression for $\rho_{DE}(a)$. Such an approach has been made popular by the search for unified DE models made out by a single fluid working as both dark matter and DE. As an example, we consider here the Hobbit model \cite{cardone04}  with

\begin{equation}
f_{DE}(a)= f_{Hob}(a) =\left ( \frac{1 + s/a}{1 + s} \right )^{\beta - \alpha} \left [ \frac{1 + (b/a)^{\alpha}}{1 + b^{\alpha}} \right ]\ 
\label{eq: rhohobvsa}
\end{equation}
with $(\alpha, \beta)$ setting the scaling in the late and early universe, respectively, and $(s, b)$ two scaling factors controlling the transition among the different regimes. Inserting (\ref{eq: rhohobvsa}) into (\ref{eq: conteq}) allows to compute $w_{DE}(a ) = w_{Hob}(a)$ which turns out to be 

\begin{equation}
w_{Hob}(a) = \frac{[(\alpha - 3) (b/a)^{\alpha} - 3] }{3 (1 + s/a) [ 1 + (b/a)^{\alpha}]} 
- \frac{ (s/a) [(\alpha - \beta + 3) + (3 - \beta) (b/a)^{\alpha}]}{3 (1 + s/a) [ 1 + (b/a)^{\alpha}]} \ .
\label{eq: whob}
\end{equation}
It is only a matter of algebra to get the following relevant asymptotic behaviours\,:

$$
\lim_{a \rightarrow 0}{w_{Hob}(a)} = \frac{\beta - 3}{3}  \ ,
$$

\begin{equation}
\lim_{a \rightarrow 1}{w_{Hob}(a)} = \frac{1}{1 + s} \left [ -1 + \frac{(\beta - 3) s}{3} + \frac{\alpha (b^{\alpha} - s)}{3 (1 + b^{\alpha})} \right ] 
\simeq -1 + \frac{\alpha b^{\alpha}}{3(1 + b^{\alpha})} \ ,
\end{equation}

\begin{displaymath}
\lim_{a \rightarrow \infty}{w_{Hob}(a)} = -1  \ ,
\end{displaymath}
where, in the second row, we have used $\alpha, \beta \ge 0$ and $s << b < 1$. These expressions show that, in the early universe, the Hobbit EoS can mimic both radiation $(\beta = 4)$ and matter ($\beta = 3$) thus eliminating the problem of early DE. At the present time, one can tailor $(\alpha, b)$ so that the EoS is close to the $\Lambda$ value $(w = -1)$ suggesting that the model is able to give an accelerated expansion. Finally, in the future, the Hobbit fluid reduces to the cosmological constant no matter the values of the model parameters. This nice behaviour motivated some of us to introduce the Hobbit model as a unified DE scenario. However, we will rather consider it as a phenomenological DE model thus also taking a dark matter contribution in Eq.(\ref{friedmann}). We let the data decide whether the matter can be reduced to the baryons contribution only.

\section{Constraining model parameters}

Both the BA and the Hobbit models are characterized by a set of parameters that can be settled in such a way to provide a cosmic scenario as close as possible to the observed one. On the one hand, the BA model may be naively matched to observation forcing $(w_0, w_a) = (-1, 0)$ so that the concordance $\Lambda$CDM scenario is recovered. On the other hand, this is not possible for the Hobbit model which has been contrasted to then available data in \cite{cardone04}, but forcing $(\alpha, \beta) = (3, 4)$ to reduce the number of parameters. In order to be more general, we therefore constrain both model parameters by fitting them to a wide dataset. We briefly describe below the data used and the Bayesian fitting procedure adopted and then present the results relevant for the later discussion.

\subsection{Data and fitting method}

As a first observable, we rely on the SNeIa Hubble diagram as traced by the JLA sample \cite{SNeIa-5}. In order to save time without loss of precision, we use the compressed distance modulus sample thus defining the likelihood function as

\begin{displaymath}
{\cal{L}}_{SNeIa}({\bf p}) =  \frac{\exp{[-\chi^2_{\mu}({\bf p})/2]}}{(2 \pi)^{{\cal{N}}_{\mu}/2} |{\bf C}_{\mu}|^{1/2}} 
\end{displaymath}
with ${\cal{N}}_{\mu}$ the number of data, 

\begin{equation}
\chi^2_{\mu}({\bf p}) = \Delta {\cal{D}}_{\mu}^{T}({\bf p}) {\bf C}_{\mu}^{-1} \Delta {\cal{D}}_{\mu}({\bf p}) \ ,
\label{eq: chisqsneia}
\end{equation}
$\Delta {\cal{D}}_{\mu}({\bf p}) = {\cal{D}}_{\mu}^{obs} - {\cal{D}}_{\mu}^{th}(\bf p)$ the vector with the difference between the observed and predicted distance moduli and ${\bf C}_{\mu}$ the covariance matrix. We remind that the distance modulus is related to the model parameters as 

\begin{displaymath}
\mu(z, {\bf p}) = 25 + 5 \log{d_L(z, {\bf p}}) + {\cal{M}}_{SNeIa}
\end{displaymath}
being

\begin{displaymath}
d_L(z, {\bf p}) = \frac{c \ (1 + z)}{H_0} \int_{0}^{z}{\frac{dz^{\prime}}{E(z^{\prime}, {\bf p})}}
\end{displaymath}
the luminosity distance and ${\cal{M}}_{SNeIa}$ a nuisance parameter (accounting for difference between the adopted zero-point of the distance scale) which we marginalize over.

The SNeIa Hubble diagram is based on the use of a standard candle. On the contrary, Baryon Acoustic Oscillations (BAO) provide a standard ruler thus offering a further probe of the background expansion. We rely here on different datasets, the lowest redshift one being at $z = 0.106$ from the 6dFGS \cite{6dFGS} team. The likelihood function is simply 

\begin{eqnarray}
{\cal{L}}_{6dFGS}({\bf p}) & = & \frac{1}{\sqrt{2 \pi} \sigma_{6dFGS}} \\
 &  \times & \exp{\left \{ - \frac{1}{2} 
\left [ \frac{d_z^{6dFGS} - d_{z}(z = 0.106, {\bf p})}{\sigma_{6dFGS}} \right ]^2 \right \}} \ , \nonumber
\label{eq: like6dFGS}
\end{eqnarray}
with $(d_z^{6dFGS}, \sigma_{6dFGS}) = (0.336, 0.015)$ and 

\begin{displaymath}
d_z(z, {\bf p}) = r_s(z_d, {\bf p})/D_V(z, {\bf p})
\end{displaymath}
being $z_d$ the drag redshift (approximated as in \cite{EH98}),

\begin{displaymath}
r_s(z, {\bf p}) = \frac{c}{H_0} \int_{z}^{\infty}{\frac{dz^{\prime}}{E(z^{\prime}, {\bf p}) 
\sqrt{3 [1 + \bar{R}_b/(1 + z^{\prime})]}}}
\end{displaymath}
the comoving sound horizon at redshift $z$ with $\bar{R}_b = 31500 \Omega_b h^2 (T_{CMBR}/2.7)^{-4}$, and 

\begin{displaymath}
D_V(z, {\bf p}) = \frac{c}{H_0} \left \{ \frac{1}{E(z, {\bf p})} 
\left [ \int_{0}^{z}{\frac{dz^{\prime}}{E(z^{\prime}, {\bf p})}} \right ]^2 \right \}^{1/3} \ .
\end{displaymath}
Moving to larger redshift, we use the results of the SDSS DR7 team to measure $d_z(z)$ at $z = (0.20, 0.35)$ \cite{DR7}. Since these data are correlated, the likelihood function now is

\begin{displaymath}
{\cal{L}}_{DR7}({\bf p}) =  \frac{\exp{[-\chi^2_{DR7}({\bf p})/2]}}{2 \pi |{\bf C}_{DR7}|^{1/2}} 
\end{displaymath}
with  

\begin{equation}
\chi^2_{DR7}({\bf p}) = \Delta d_z^{T}({\bf p}) {\bf C}_{DR7}^{-1} \Delta d_{z}({\bf p}) \ ,
\label{eq: chisqdr7}
\end{equation}
and $\Delta d_{z}({\bf p}) = d_{z}^{obs} - d_{z}(\bf p)$. The observed values $d_z^{obs}$ and the covariance matrix ${\bf C}_{DR7}$ are given in \cite{DR7}.

Both the BOSS and WiggleZ collaborations have measured BAO at still larger $z$ observing galaxies over a partially overlapping survey area. In \cite{BW}, the combined dataset has been used to measure 

\begin{displaymath}
\tilde{D}_V(z, {\bf p}) = D_V(z, {\bf p}) r_{s}^{fid}/r_s(z_d, {\bf p})
\end{displaymath}
where $r_s^{fid}$ is a fiducial value adopted in the analysis of the single samples. We include these measurements in our analysis defining the likelihood function

\begin{displaymath}
{\cal{L}}_{BW}({\bf p}) =  \frac{\exp{[-\chi^2_{BW}({\bf p})/2]}}{(2 \pi)^2 |{\bf C}_{BW}|^{1/2}} 
\end{displaymath}
with  

\begin{equation}
\chi^2_{BW}({\bf p}) = \Delta \tilde{D}_V^{T}({\bf p}) {\bf C}_{BW}^{-1} \Delta \tilde{D}_V({\bf p}) \ ,
\label{eq: chisqbw}
\end{equation}
being $\Delta \tilde{D}_V({\bf p})$ the usual difference vector. We refer the reader to \cite{BW} for the observed $\tilde{D}_V$ at $z = (0.57, 0.73)$, the covariance matrix ${\bf C}_{BW}$, and the $r_s^{fid}$ values. Note that all the BAO likelihood functions also depend on the baryon density parameter $\Omega_b$ which we marginalize over.

The dataset described above probe the late universe so that it is interesting to add observables related to the early epochs. These are naturally provided by the CMB distance priors which summarize the information content of the Planck13 data \cite{Planck13}. The three quantities of interest are the acoustic scale

\begin{displaymath}
\ell_A({\bf p}) = \frac{\pi d_L(z_{LSS}, {\bf p})}{(1 + z_{LSS}) r_s(z_{LSS}, {\bf p})} 
\end{displaymath}
with $z_{LSS}$ the last scattering surface redshift estimated as in \cite{EH98}, the shift parameter

\begin{displaymath}
{\cal{R}}({\bf p}) = \frac{\sqrt{\Omega_M} H_0}{c} \frac{d_L(z_{LSS}, {\bf p})}{1 + z_{LSS}} \ ,
\end{displaymath}
and the physical baryon density $\omega_b = \Omega_b h^2$. The CMB distance priors likelihood function is then defined as

\begin{displaymath}
{\cal{L}}_{CMB}({\bf p}) =  \frac{\exp{[-\chi^2_{CMB}({\bf p})/2]}}{2 \pi |{\bf C}_{DR7}|^{1/2}} 
\end{displaymath}
with  

\begin{equation}
\chi^2_{CMB}({\bf p}) = \Delta_{CMB}^{T}({\bf p}) {\bf C}_{CMB}^{-1} \Delta_{CMB}({\bf p}) \ ,
\label{eq: chisqcmb}
\end{equation}
with

\begin{displaymath}
\Delta_{CMB}({\bf p}) = \left [
\begin{array}{l}
\ell_A({\bf p}) - 301.57 \\
{\cal{R}}({\bf p}) - 1.7407 \\
\Omega_b h^2 - 0.02228 \\
\end{array}
\right ] \ ,
\end{displaymath}
and the covariance matrix ${\bf C}_{CMB}$ given in \cite{Planck13}.

All the data we have considered so far depend on the integrated Hubble expansion rate. It is therefore interesting to complement them with direct measurements of $H(z)$ itself. We use the compilation in \cite{Duan16} which comes from an heterogenous sample of data obtained with different techniques. This ensures us that the covariance matrix is diagonal so that the corresponding likelihood has the typical Gaussian form ${\cal{L}}_{H}({\bf p}) \propto \exp{[-\chi_H^2({\bf p})/2 ]}$ with

\begin{equation}
\chi_H^2({\bf p}) = \sum_{i = 1}^{{\cal{N}}_H}{\left [ \frac{H_{obs}(z_i)  - H_0 E(z_i, {\bf p})}{\sigma_i} \right ]^2} \ .
\label{eq: defchih}
\end{equation}
Finally, to set the zero-point, we also add a Gaussian prior on $h$ (i.e., $H_0$ in units of $100 \ {\rm km/s/Mpc}$) setting $h_{obs} = 0.706 \pm 0.033$ as in \cite{E13}.

In order to perform a combined analysis, we maximize the full likelihood simply defined as the product of the single terms introduced above. To this end, we use a standard Markov Chain Monte Carlo (MCMC) method to sample the parameter space running six chains of equal length and keep adding points until convergence is achieved. The merged chain (after burn in cut and thinning) is then used to infer median and confidence ranges for each parameter.

\subsection{Results}

\begin{figure*}
\centering
\includegraphics[width=6.0cm]{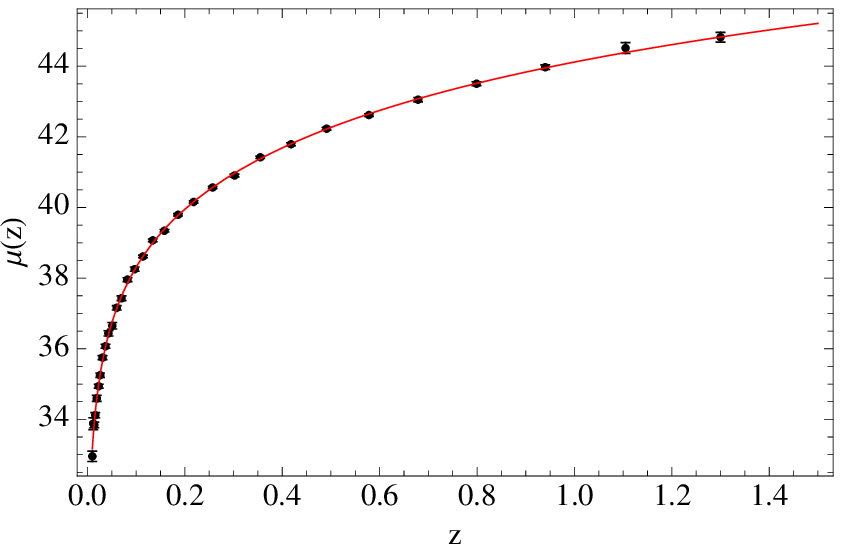}
\includegraphics[width=6.0cm]{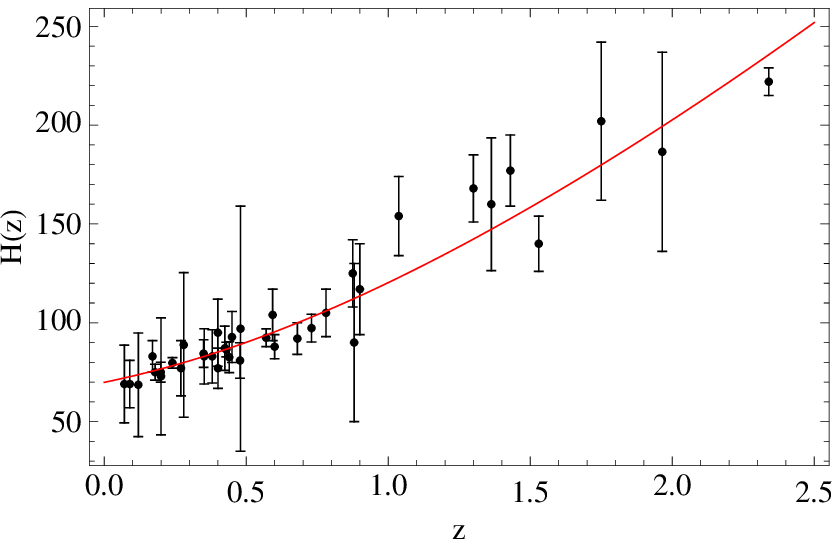} \\
\includegraphics[width=6.0cm]{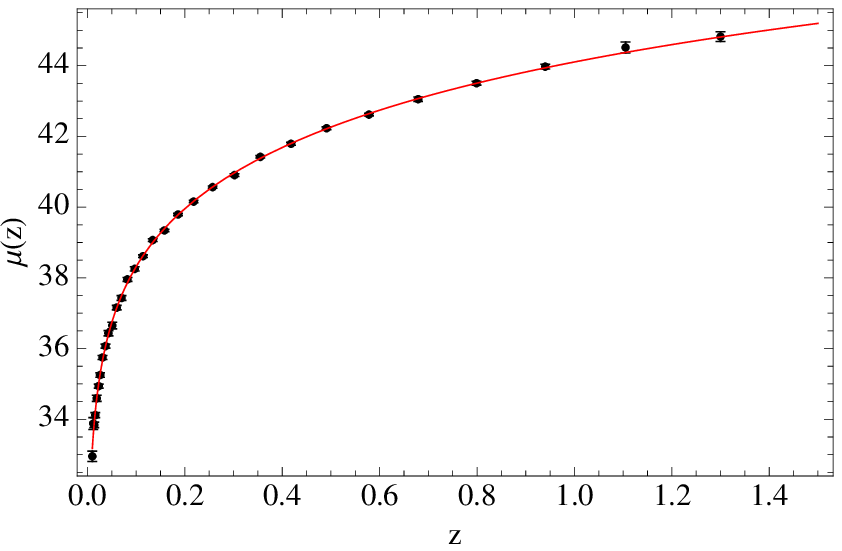}
\includegraphics[width=6.0cm]{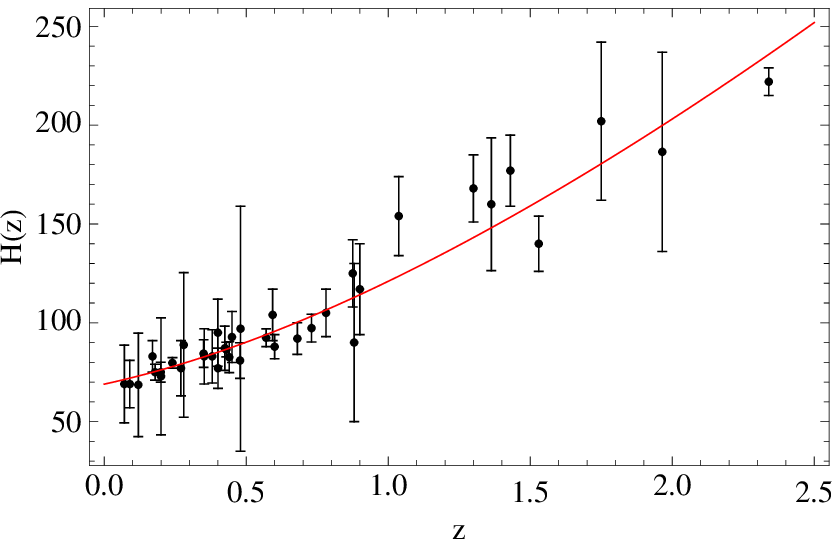} \\
\caption{Best fit models superposed to the SNeIa and $H(z)$ data for both the BA (top) and Hobbit (bottom) scenarios.}
\label{fig: bestfit}
\end{figure*}

We fit both the BA and Hobbit models to the large dataset described in the previous paragraph. As can be seen from Fig.\,\ref{fig: bestfit}, both models work quite well at reproducing the SNeIa Hubble diagram and the $H(z)$ data. In order to get these plots, we set the model parameters to their best fit values, namely

\begin{displaymath}
\Omega_M = 0.290 \ , \ w_0 = -1.010 \ ,\ w_a = -0.076 \ , \ h = 0.699 \ , 
\end{displaymath}
for the BA model, and 

\begin{displaymath}
\Omega_M = 0.294 \ , \ \alpha = 0.093 \ , \ \beta = 0.443 \ , \ 
b = 0.540 \ , \ s = 2.143 \times 10^{-4} \ , \ h =  0.689 \ , 
\end{displaymath}
for the Hobbit case. Note that the radiation density parameter has not been allowed to vary in the fit, but we rather set

\begin{displaymath}
\Omega_R = \omega_{\gamma} h^{-2} (1 + 0.2271 N_{eff})
\end{displaymath}
with $\omega_\gamma = 2.475 \times 10^{-5}$ and $N_{eff} = 3.04$ the effective number of massless neutrinos. While the best fit values for the BA model are expected since they are close to the $\Lambda$CDM limit $(w_0, w_a) = (-1, 0)$, the $(\alpha, \beta)$ values for the Hobbit model are somewhat surprising being quite different from the case $(\alpha, \beta) = (3, 4)$ which we have identified as particularly interesting. Actually, such a small value for $\alpha$ can be easily explained going back to the approximated formula for $w_{Hob}(a = 1)$. Inserting the best fit $(\alpha, b)$ values quoted above gives $w_{Hob}(a = 1) = -0.982$ which is very close to the $\Lambda$CDM concordance scenario. The best fit $\beta$ is then set in such a way that $w_{Hob}(a)$ asymptotes to $w_{DE} = -0.85$ which is still close to the $\Lambda$CDM $w = -1$ value. Put in other words, both the BA and Hobbit best fit models try to mimic as close as possible the concordance $\Lambda$CDM scenario which is somewhat expected given the observational success of the cosmological constant.

As a further confirmation of this qualitative discussion, we stress that both models perform equally well to reproduce not only the SNeIa and $H(z)$ data, but also the BAO clustering constraints and the Planck13 priors. As a consequence, the overall likelihood turns out to be almost equal for the two models although this argues in favor of the BA one which has a smaller number of parameters. It is worth noting, however, that the two models have a distinct evolutionary history with $w_{BA}(a)$ remaining almost constant all over the universe expansion due to the very small $w_a$ value. On the contrary, $w_{Hob}(a)$ presents a non negligible evolution moving from $w_{Hob} = -0.98$ today to $w_{Hob} = -0.85$ in the radiation era.

Although we do not need them in the following, we also report below the constraints (median and $68\%$ confidence values) on the model parameters. For BA we find 

\begin{displaymath}
\Omega_M = 0.256_{-0.087}^{+0.058} \ \ ,  \ \ 
w_0 = -0.947_{-0.453}^{+0.339}  \ \ , \ \ 
w_a  = -0.013_{-0.149}^{+0.143} \ \ , \ 
h = 0.707_{-0.060}^{+0.097} \ \ ,
\end{displaymath}  
For the Hobbit model, we have used $\eta_b = \log{(1/b)}$ and $\eta_s = \log{(1/s)}$ as parameters instead of $(b, s)$ since they allow to explore a wider range. We then find

\begin{displaymath}
\Omega_M = 0.223_{-0.100}^{+0.084} \ \ , \ \ 
h = 0.708_{-0.050}^{+0.120} \ \ ,
\end{displaymath}

\begin{displaymath}
\alpha = 0.508_{-0.307}^{+0.692} \ \ , \ \
\beta = 1.014_{-0.573}^{+1.280} \ \ , 
\end{displaymath}

\begin{displaymath}
\eta_b = 0.158_{-0.076}^{+0.135} \ \ , \ \ 
\eta_s = 3.468_{-0.428}^{+0.356} \ \ .
\end{displaymath}
We note that both the matter density parameter $\Omega_M$ and the present day Hubble constant $h$ are well consistent between the two models. While the BA parameters are severely constrained to lie close to their $\Lambda$CDM counterparts, some more discussion is needed to interpret the results for the Hobbit model. Indeed, the median $(\alpha, \beta)$ values are quite different from the best fit ones which are outside the $68\%$ CL (but within the not reported $95\%$ ones). Such an unusual behaviour is a consequence of two effects. First, strong degeneracies among the model parameters can shift the best fit value from the median ones. Indeed, the best fit parameters are chosen to maximize the likelihood, while the median ones characterize the probability distribution function of each parameter after taking into account the fit to the data. The two quantities are different so that it is not unusual to find discrepant values when a large parameter space is constrained by a limited amount of data. Second, most of the data probe the late universe with only Planck priors tracing the early one and no data at all covering the matter dominated region where the Hobbit EoS can change from cosmological constant\,-\,like to dust\,-\,like. This prevents to put significant constraints on both $\beta$ and $\eta_b$ thus propagating the uncertainty to other parameters too. Motivated by these considerations, we do not consider troublesome the apparent discrepancy among best fit and median $(\alpha, \beta)$ values. However, since in the following we will be mainly interested in showing how thermodynamics can discriminate between models fitting equally well the same dataset, we will set both the BA and Hobbit model parameters to their best fit values.

\section{Thermodynamic analysis}\label{sec:thermodynamics}

It is customary to consider a DE model as well motivated if it passes the matching with observational data with green lights. However,  the realm of observationally viable models is always increasing. Lacking an underlying conclusive theory, it is worth addressing the problem from a different point of view. Thermodynamics offers such a distinct look. Indeed, we aim here at investigating whether different models fitting equally well the same data can be distinguished on the basis of their thermodynamical properties. Put in other words, we wonder whether thermodynamics can alleviate the degeneracy surviving the data fitting test. 

In a cosmological framework, we actually deal with the Generalized Second Law (GSL) of thermodynamics since, for system with some horizon, the entropy of the horizon itself must be added to those of the fluids within it. In a sense, this arises in a natural way from the consideration that the horizon prevents the observer to see what lies beyond it.

As already seen elsewhere \cite{radicella12}, this approach is theoretically well defined in cosmology and translates, within the FRW framework, in the requirements $S^{\prime}(a) \geq 0$ and, in the far future, as a consequence of the generalized second law of thermodynamics, $S^{\prime \prime}(a) \le 0$, where the prime denotes the derivative with respect to $a$. In such a case, an apparent horizon is naturally defined as well as the relative entropy, which depends on the surface, in analogy with BHs; on the other hand standard thermodynamical definitions of entropy can be applied to the matter\,-\,energy components. The appropriate boundary in this context has been shown to be the apparent horizon of FRW universes, which always exists, something that is not generally true for the particle horizon and the future event horizon \cite{wang06, cai09}.

Let us first compute the different contribution to the total entropy starting from the horizon one. To this end, we take the apparent horizon, i.e., the marginally\,-\,trapped surface with vanishing expansion, since - by contrast to other possible choices - the laws of thermodynamics are fulfilled on it \cite{wang06}. Writing the line element as

\begin{displaymath}
ds^2 = h_{ab} dx^a dx^b + \tilde{r}^2 d\Omega^2 \ ,
\end{displaymath}
the apparent horizon is defined by the condition $h^{ab}\partial_a\tilde{r} \partial_b\tilde{r} =0$, being $\tilde{r} = a(t) r$ and  $h_{ab} = \text{diag}(-c^2,a(t)^2)$. Its radius turn out to be $\tilde{r}_A = c/H(a)$ so that the area is naively ${\cal{A}} = 4 \pi \tilde{r}_A^2$. The entropy is simply proportional to ${\cal{A}}$ so that we get

\begin{equation}
S_H = \frac{k_B}{4} \frac{{\cal{A}}}{\ell_{Pl}}
\label{eq: sh}
\end{equation}
with $k_B$ the Boltzmann constant and $\ell_{Pl}$ the Planck length. It is worth noting that $S_H$ is defined in terms of the Hubble expansion rate $H(a)$ and hence is determined by the properties (abundance and EoS) of the fluids contributing to the energy budget. Actually, in order to verify thermodynamical prescriptions, rather than the entropy itself, we are interested in its derivatives with respect to the scale factor $a$. 

Restoring physical constants and using the Friedmann and continuity equations, we finally get

\begin{equation}
\tilde{S}^{\prime}_H(a) = \frac{4 \pi}{a E^2(a)} \left [ 1 + \sum_{i}{w_i(a) \Omega_i(a)} \right ]
\label{eq: spH}
\end{equation}
where the sum is over the fluids in the universe and hereafter we will use the dimensionless entropies defined as

\begin{displaymath}
\tilde{S}_X(a) = \left ( \frac{3 k_B c^5}{4 G \hbar H_0^2} \right )^{-1} S_X(a) \ .
\end{displaymath}
Making use of the Eq.(\ref{continuity}), we then get the following result for the second derivative of the apparent horizon entropy

\begin{equation}
\frac{\tilde{S}^{\prime \prime}_H(a)}{3/a^2 E^2(a)} \propto  
\sum_{i}{a w^{\prime}_i(a) \Omega_i(a)} + 6 \left [ \sum_{i}{w_i(a) \Omega_i(a)} \right ]^2 
+ \sum_{i}{w_i(a) [5 - 3 w_i(a)] \Omega_i(a)}  + 2 \ .
\label{eq: sppH}
\end{equation}
The approach to the thermodynamical equilibrium of the horizon $(\tilde{S}^{\prime \prime}_H = 0)$ is then driven by the EoS of the fluid components, determining their abundance evolution, and its derivative. 

To compute the entropy of each fluids within the horizon, we must first determine how their temperature evolves. This readily follows from the Gibbs equation and the condition that $dS_i$ is a differential. We then get

\begin{equation}
\frac{d\ln{T_i(a)}}{d\ln{a}} = -3 w_i(a) 
\label{Eq: tvsa}
\end{equation}
which can be trivially integrated given the fluid EoS. For radiation $(w = 1/3)$, we get 

\begin{equation}
\tilde{T}_R(a) = T_R(a)/T_{0R} = a
\label{eq: tildetr}
\end{equation}
where the present day value may be set equal to the CMBR one, i.e. $T_{0R} = T_{CMBR} = 2.725 \ {\rm K}$. For the DE, we get two different expressions depending on which EoS is used. Setting $w_{DE}(a) = w_{BA}(a)$ gives

\begin{equation}
\tilde{T}_{DE}(a) = a^{-3(w_0 + w_a)} (1 - 2 a + 2 a^2)^{3 w_a/2} \ ,
\label{Eq: tvsacpl}
\end{equation}
while using the Hobbit DE we get 

\begin{equation}
\tilde{T}_{DE}(a) = \left ( \frac{1 + s/a}{1 + s} \right )^{\beta - \alpha} \left [ \frac{1 + (b/a)^{\alpha}}{1 + b^{\alpha}} \right ] a^3 \ .
\label{Eq: tvsahob}
\end{equation}
For both scenarios, $\tilde{T}_{DE}(a) = T_{DE}(a)/T_{0DE}$ and the present day temperature of the DE is unknown so that we set it as $T_{0DE} = \tau_{DE} H_0 \hbar/k_B$ leaving $\tau_{DE}$ as an additional parameter.

We can now rely on the Euler equation $ T s =\rho(1 + w)$ with $s$ the entropy density to first compute the entropy of the $i$\,-\,th fluid and then its derivative with respect to $a$. Going to dimensionless quantity, we find

\begin{equation}
\tilde{S}^{\prime}_i(a) = \left [ \frac{H_0 \hbar}{ k_B T_{0i} \tilde{T}_{i}(a)} \right ]
\frac{\Omega_i(a) [1 + w_i(a)]}{a E(a)}  
\left [ 1 + 3 \sum_{j}{w_j(a) \Omega_j(a)} \right ] \ .
\label{eq: fluidentropy}
\end{equation}
A similar derivation can not be used for the matter term since its temperature is identically null all along the cosmic history. We can instead use $S_M = k_B V_{H} n$, where $V_{H}$ is the volume of the apparent horizon and $n$ the number density of dust particles. Differentiating with respect to $a$ finally gives

\begin{equation}
\tilde{S}^{\prime}_M(a) = \left ( \frac{H_0 \hbar}{k_B {\cal{T}}_{0M}} \right )
\frac{1}{a^4 E^3(a)}  
\left [ 1 + 3 \sum_{i}{w_i(a) \Omega_i(a)} \right ] \ ,
\label{eq: spM}
\end{equation}
where we have defined

\begin{equation}
{\cal{T}}_{0M} = \frac{3 c^2 H_0^2}{8 \pi G k_B n_0} \ \ , 
\label{eq: deft0m}
\end{equation}
with $n_0 \sim 10^{-6} \ {\rm m}^{-3}$ the present day dust number density\footnote{This is the typical number density of dust particles in the space, but its value is actually irrelevant for following considerations.}. We stress that, although with the same dimension, ${\cal{T}}_{0M}$ is not the matter temperature, but only a convenient quantity introduced to give $\tilde{S}^{\prime}_M(a)$ the same form as the radiation and DE terms. 

The total entropy is the sum of the horizon entropy and those relative to each fluid component. Total first derivative of entropy, $\tilde{S}^{\prime}_{tot}$, can be conveniently written as

\begin{equation}
\tilde{S}^{\prime}_{tot}(a)  = \frac{H_0 \hbar}{k_B T_{0R}}  
\frac{{\cal{W}}_2(a) w_{DE}^2(a) + {\cal{W}}_1(a) w_{DE}(a) + {\cal{W}}_0(a)}{a E(a)}
\label{eq: sptot}
\end{equation}
with

\begin{equation}
{\cal{W}}_0(a) = \frac{4 \pi k_B T_{0R}}{H_0 \hbar E(a)} \left [ 1 + \frac{\Omega_R(a)}{3} \right ]
+ \left [ \frac{4 \Omega_R(a)}{3 \tilde{T}_R(a)} + \frac{T_{0R}}{{\cal{T}}_{0M}} \frac{1}{a^3 E^2(a)} + 
\frac{T_{0R}}{T_{0DE}} \frac{\Omega_{DE}(a)}{\tilde{T}_{DE}(a)} \right ] 
\left [ 1 + \Omega_R(a) \right ]  \ , 
\label{eq: defw0}
\end{equation}

\begin{equation}
\frac{{\cal{W}}_1(a)}{\Omega_{DE}(a)} = \frac{4 \Omega_R(a)}{\tilde{T}_R(a)} + \frac{T_{0R}}{{\cal{T}}_{0M}} \frac{3}{a^3 E^2(a)} 
+ \frac{4 \pi k_B T_{0R}}{H_0 \hbar E(a)}
+ \frac{T_{0R}}{T_{0DE}} \frac{1 + \Omega_R(a) + 3 \Omega_{DE}(a)}{\tilde{T}_{DE}(a)} \ ,
\label{Eq: defw1}
\end{equation}

\begin{equation}
{\cal{W}}_2(a) = \frac{3 \Omega_{DE}^2(a)}{\tilde{T}_{DE}(a)} \frac{T_{0R}}{T_{0DE}} \ .
\label{eq: defw2}
\end{equation} 
having used $w_{R}(a) = 1/3$. It is instructive to estimate the order of magnitude of the terms entering Eqs.(\ref{eq: defw0})\,-\,(\ref{eq: defw2}). To this end, we insert the constants with the correct units in the above definitions to get

\begin{displaymath}
\frac{4 \pi k_B T_{0R}}{H_0 \hbar} = \frac{1.38 \times 10^{30}}{h} \ , 
\end{displaymath}
\begin{displaymath}
\frac{T_{0R}}{{\cal{T}}_{0M}} = \frac{1.70 \times 10^{-31}}{h^2} \ , 
\end{displaymath}
\begin{displaymath}
\frac{T_{0R}}{T_{0DE}} = \frac{1.01 \times 10^{29}}{\tau_{DE} h} \ ,
\end{displaymath}
It is therefore evident that, unless in the very early universe (i.e., $a \rightarrow 0$), the matter related terms in Eqs.(\ref{eq: defw0})\,-\,(\ref{eq: defw2}) are always negligible thus explaining why we do not care about the exact $n_0$ value.

Finally, it is immediate to get the second derivative of the entropy with respect to the scale factor $a$. Differentiating Eq.(\ref{eq: sptot}), we finally get

\begin{eqnarray}
\tilde{S}^{\prime \prime}_{tot}(a) & = & 
\frac{2 {\cal{W}}_2(a) w_{DE}^{\prime}(a) + {\cal{W}}_1(a) w_{DE}^{\prime}(a)}{a E(a)} \nonumber \\
 & + & \frac{{\cal{W}}^{\prime}_2(a) w_{DE}^2(a) + {\cal{W}}^{\prime}_1(a) w_{DE}(a) + {\cal{W}}^{\prime}_0(a)}{a E(a)} \nonumber \\
 & - & \frac{\tilde{S}^{\prime}_{tot}(a)}{a} \left [ 1 + \frac{d\ln{E(a)}}{d\ln{a}} \right ] \ ,
\label{eq: spptot}
\end{eqnarray}
where we do not report the derivatives of the different terms for sake of shortness.

Eqs.(\ref{eq: sptot}) and (\ref{eq: spptot}) show that the variation of the total entropy depend on the evolution of the density parameter and the EoS of all the fluids contributing to the full energy budget. Actually, the two by far dominant contributions are the horizon and DE ones with the latter playing a crucial role in the fulfillment of the GSL. Indeed, being the only fluid with negative EoS, it is also the only one which can drive $\tilde{S}^{\prime}_{tot}(a)$ towards negative values. It is the balance between the DE and horizon terms which will finally determine whether the entropy decrease or increase during the universe history. 

\section{Thermodynamic constraints}

As yet said before, given the strong connection between gravity and thermodynamics \cite{bekenstein}, we suppose that the Universe behaves as an isolated thermodynamic systems so that it approaches to a state of maximum entropy in the long run. As a consequence, in order for a cosmological model to be viable from a thermodynamical point of view, the GSL must be fulfilled so that two following conditions must be passed

\begin{equation}
\tilde{S}^{\prime}_{tot}(a) \ge 0 \ , 
\label{eq: spcons}
\end{equation}

\begin{equation}
\lim_{a \rightarrow \infty}{\tilde{S}^{\prime \prime}_{tot}(a)} \le 0 \ .  
\label{eq: sppcons}
\end{equation}
Since the entropy can never decrease, Eq.(\ref{eq: spcons}) must hold over all the universe history from the beginning to its end, i.e. for $0 \le a \le a_{max}$, with $a_{max}$ a very large value. However, both DE models we are considering have been proposed as phenomenological descriptions of this component which we expect to be valid over a limited range in redshift. We will therefore impose Eq.(\ref{eq: spcons}) over the range $(a_{min}, a_{max})$ with $a_{min} = (1 + z_{LSS})^{-1} \sim 0.001$ and $a_{max} = (1 + z_{min})^{-1} \simeq 1000$ where $z_{LSS} \sim 1090$ and $z_{min} = -0.999$ are the redshifts of the last scattering surface redshift and of distant point in the future close enough to the asymptotic limit $a \rightarrow \infty$.

\begin{figure*}
\centering
\includegraphics[width=8.0cm]{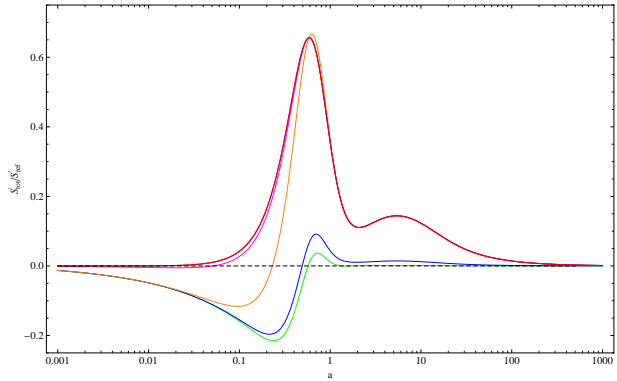}
\caption{Scaled first derivative of the entropy for the fiducial BA model. Green, blue, orange, magenta, purple, red lines refers to models with $\log{\tau_{DE}} = (-3, -2, -1, 0, 1, 2)$ and $\log{S^{\prime}_{ref}} = (3.0, 2.0, 1.0, 1.0, 1.0, 1.0)$ with $\tilde{S}^{\prime}_{ref}$ an arbitrary constant introduced to reduce the amplitude of $S^{\prime}_{tot}$ so that all curves enter in the same plot.}
\label{fig: spbalfid}
\end{figure*}

\subsection{BA model}

Let us first consider the phenomenological BA parameterization for the DE EoS. Plugging in the relevant formulae for the density parameters and temperature scaling, we can investigate whether the thermodynamical constraints are fulfilled or not.  To this end, we first set the model parameters $(\Omega_M, w_0, w_a, h)$ to their best fit values. In such a case we get the results in Fig.\,\ref{fig: spbalfid} where we plot\footnote{Since the amplitude of  $\tilde{S}^{\prime}_{tot}(a)$ changes by orders of magnitude with $\tau_{DE}$, here and in the following plots, we scale the derivatives with respect to an arbitrary chosen value which we adjust with $\tau_{DE}$ so that all the curves may be shown in the same plot. We therefore warn the reader to only look at the shape of $\tilde{S}^{\prime}_{tot}(a)$ and not to directly compare the values.} $\tilde{S}^{\prime}_{tot}(a)$ for different values of $\tau_{DE}$. 

As a remarkable result, we find that $\tilde{S}^{\prime}_{tot}(a)$ is not positive definite, but may rather become negative for a limited period of time. Put in other words, the function $\tilde{S}^{\prime}_{tot}(a)$ has at least one zero whose value depends on the present day DE temperature. Such a behaviour can be qualitatively explained as follows. As universe expands, the derivative of the DE entropy decreases until becoming negative, but the positive contribution of the horizon term becomes more and more important thus delaying the violation of the constraint (\ref{eq: spcons}). Eventually, the horizon term becomes the only one, but now the universe content is dominated by the negative pressure fluid so that the horizon entropy crosses the zero line leading to $\tilde{S}^{\prime}_{tot}(a) < 0$. The $\tau_{DE}$ parameter then controls the transition between the two regimes, while the cosmological parameters $(\Omega_M, w_0, w_a)$ determine the sign of the term in square parentheses in Eq.(\ref{eq: spH}) which controls the sign of $\tilde{S}^{\prime}_H(a)$. Note that, in the limit $(w_0, w_a) = (-1, 0)$, the horizon entropy derivative asymptotes to the null value since $\Omega_{DE}(a >> 1) \simeq 1$ and $w_{DE} = -1$. Such a condition is independent on the value of $\tau_{DE}$ so that it is not surprising that all the curves in Fig.\,\ref{fig: spbalfid} asymptotically approach an almost null value given that the fiducial $(w_0, w_a)$ parameters are indeed close to the critical ones.

\begin{figure*}
\centering
\includegraphics[width=5.0cm]{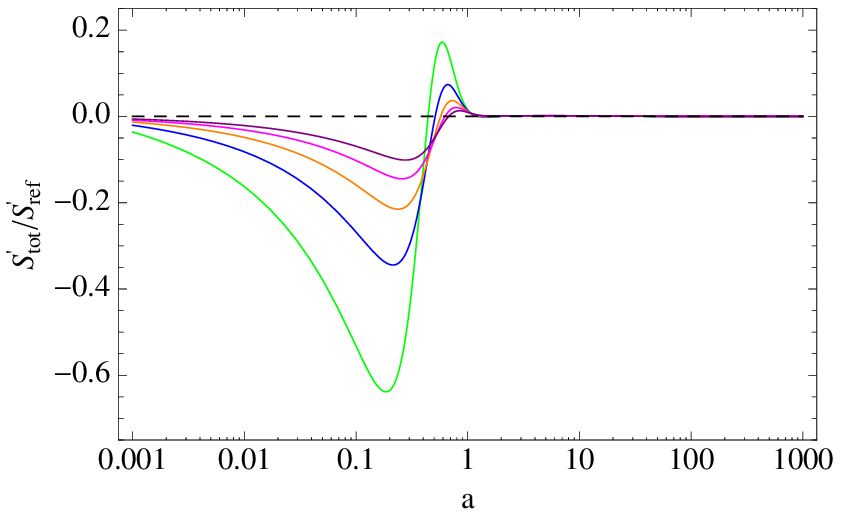}
\includegraphics[width=5.0cm]{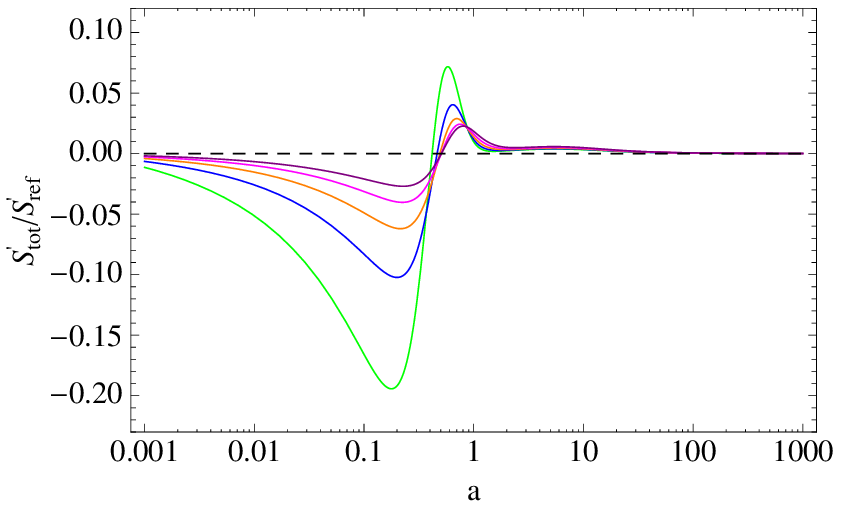}
\includegraphics[width=5.0cm]{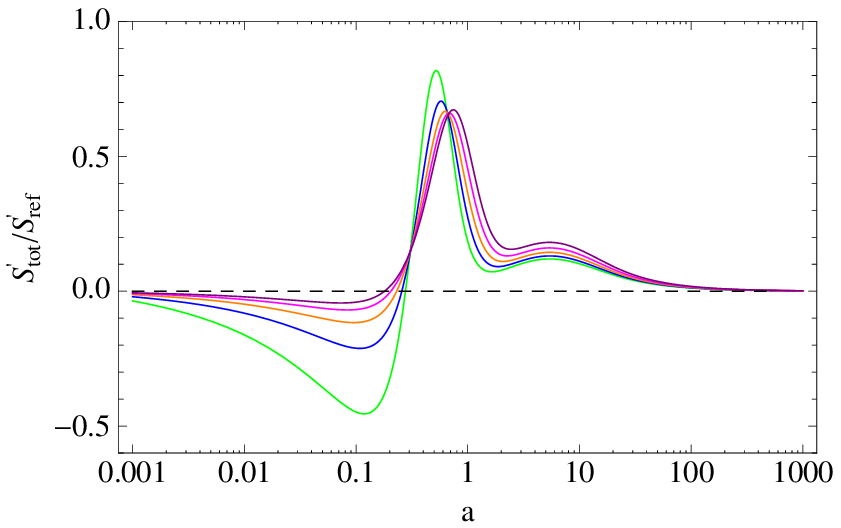} \\
\includegraphics[width=5.0cm]{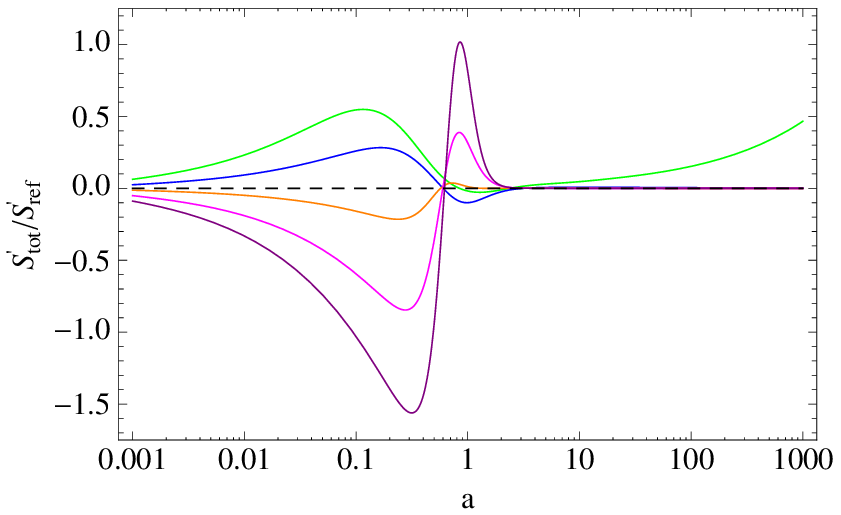}
\includegraphics[width=5.0cm]{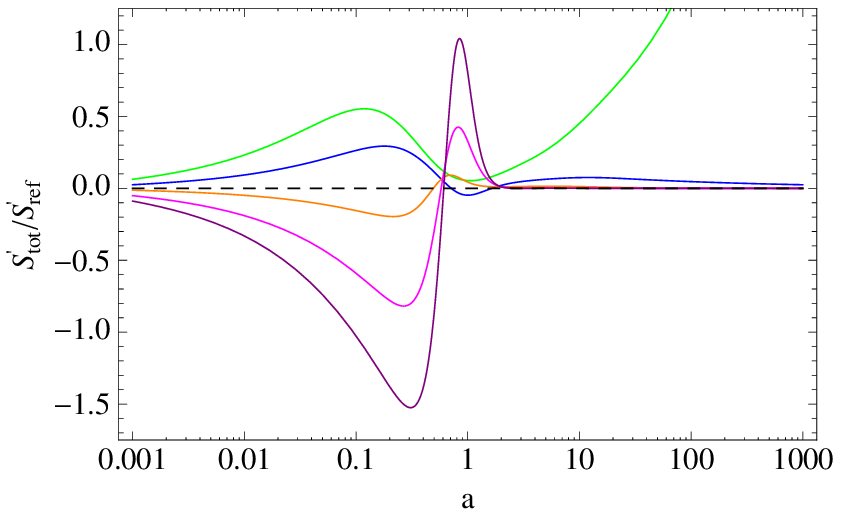}
\includegraphics[width=5.0cm]{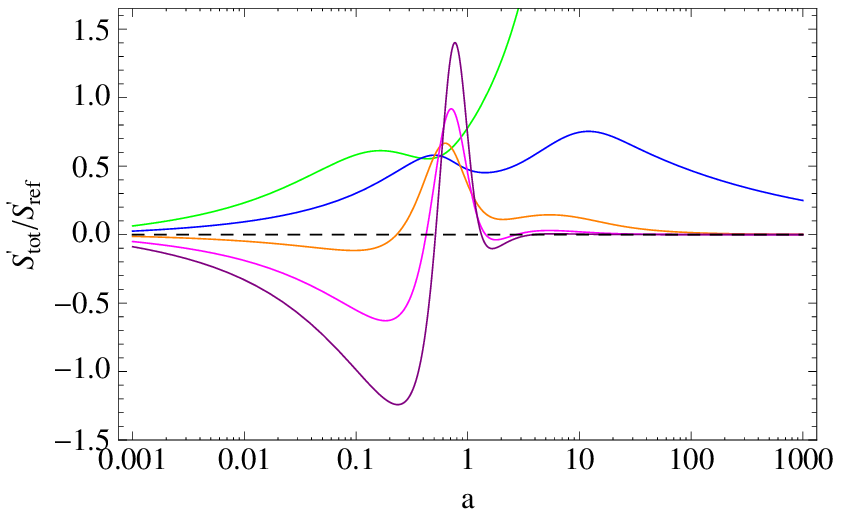} \\
\includegraphics[width=5.0cm]{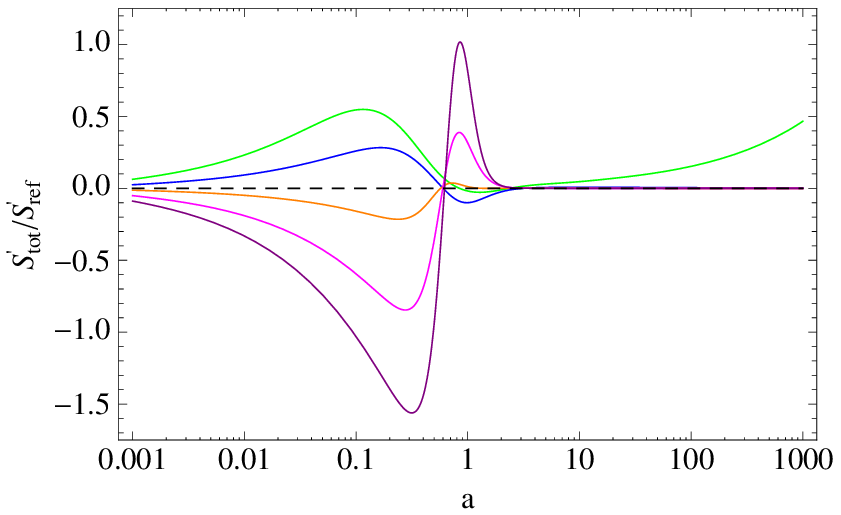}
\includegraphics[width=5.0cm]{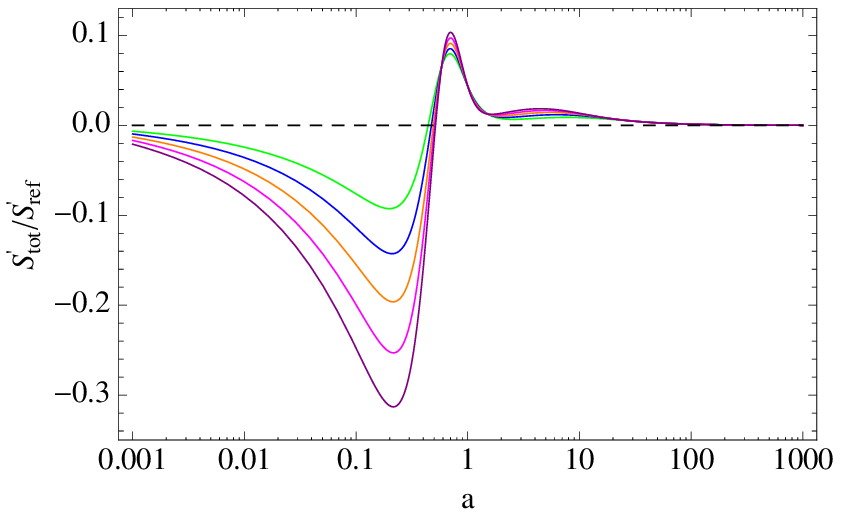}
\includegraphics[width=5.0cm]{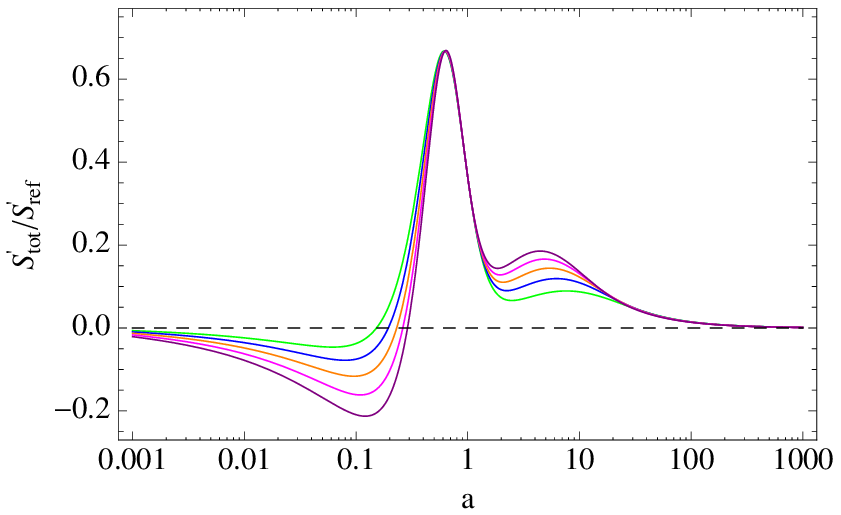} \\
\caption{Scaled first derivative of the entropy for BA models with $\log{\tau_{DE}} = (-3.0, -2.0, -1.0)$ from left to right columns. Green, blue, orange, magenta, purple lines refers to models with $p/p_{fid} = (0.50, 0.75, 1.00, 1.25, 1.50)$ with $p = (\Omega_M, w_0, w_a)$ from top to bottom rows and other parameters set to their fiducial values, being $p_{fid}$ the fiducial one for the parameter $p$. Reference values are $\log{S^{\prime}_{ref}} = (3.0, 2.0, 1.0)$ from left to right.}
\label{fig: spbalpar}
\end{figure*}

One could wonder whether such a result is specific of the fiducial model or can it be cured by changing the model parameters. In Fig.\,\ref{fig: spbalpar}, we plot $\tilde{S}^{\prime}_{tot}(a)$ changing one parameter at time and leaving the other fixed to their fiducial values for different $\tau_{DE}$. For all cases we consider, there is always a range where $\tilde{S}^{\prime}_{tot}(a)$ becomes negative so that the entropy is decreasing in that period which is unphysical. Actually, such a result is consistent with the thermodynamical analysis of DE models presented in e.g., \cite{radicella12}. As we go back in time, the DE becomes subdominant (i.e., there is no early dark energy), but its contribution to the entropy derivative is still non negligible (because of the factor $T_{0R}/{\cal{T}}_{0M}$ which makes the matter one negligible). Since $w_{BA}(a \rightarrow 0) = w_0 + w_a$, for fixed $w_a$, the larger is $w_0/w_0^{fid}$, the more the DE fluid behaves as a phantom one driving the dominant $\tilde{S}^{\prime}_{H}$ towards negative values because of the term $\Omega_{DE} w_{DE}$ in the square parenthesis. This explains why $\tilde{S}^{\prime}_{tot}(a)$ has the largest variation with $w_0$ shown in the central row of Fig.\,\ref{fig: spbalpar}, while the similar scaling one would expect for the dependence on $w_a$ is suppressed by a fiducial value which is close to zero. Note that, no matter which is the value of $\tau_{DE}$, models with small $\Omega_M$ have a negative derivative of the entropy over almost the full evolutionary history approaching then null value in the far future because of the EoS asymptotic to the cosmological constant value. This leads us to argue against them, but they are actually already excluded by the comparison with the data. A somewhat strange trend takes place for $w_0/w_{0,fid} = 0.5$ giving a divergent entropy. It is not easy to trace the origin of this unrealistic behaviour, but we nevertheless note that such values are outside the confidence range allowed by the data. \\

\begin{figure*}
\centering
\includegraphics[width=6.0cm]{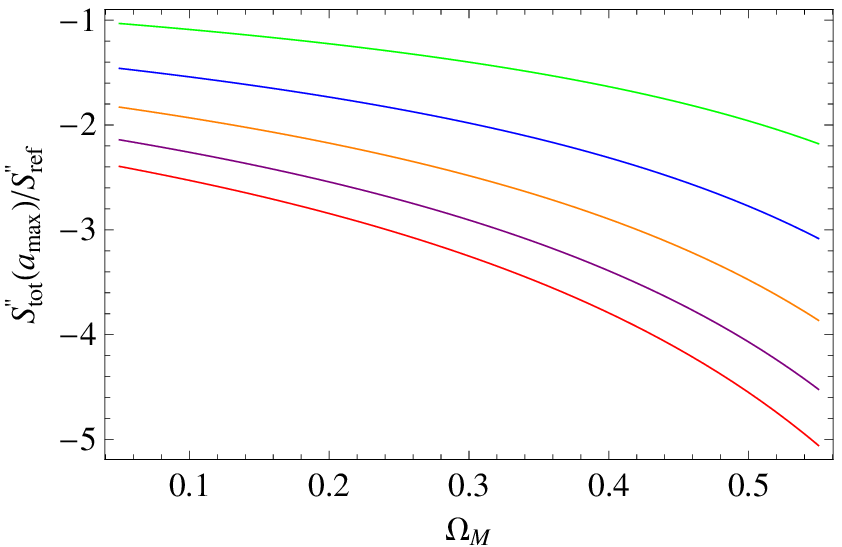}
\includegraphics[width=6.0cm]{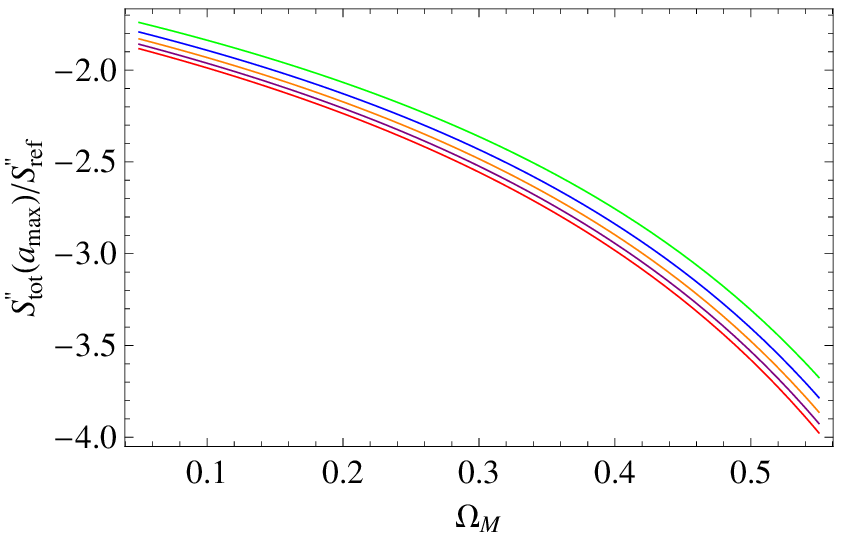} \\
\caption{Asymptotic second derivative of the entropy scaled with respect to a reference value as a function of $\Omega_M$. We set $\tau_{DE} = 1.0$ and all the parameters to their fiducial values but one with with green, blue, orange, magenta, purple lines referring to $p/p_{fid} = (0.50, 0.75, 1.00, 1.25, 1.50)$ and $p = (w_0, w_a)$ for left and right panels, respectively. We also set $\log{S^{\prime \prime}_{ref}} = (0, -3, -5, -5, -9)$ in the left panel and $\log{S^{\prime \prime}_{ref}} = -5$} in the right one.
\label{fig: sppbal}
\end{figure*}

\begin{figure*}
\centering
\includegraphics[width=8.0cm]{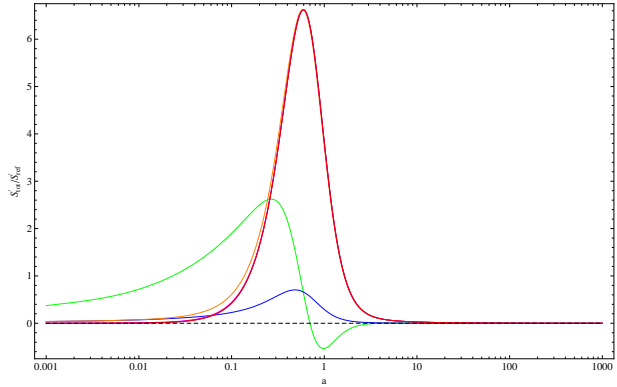}
\caption{Scaled first derivative of the entropy for the fiducial Hobbit model. Green, blue, orange, magenta, purple, red lines refers to models with $\log{\tau_{DE}} = (-3, -2, -1, 0, 1, 2)$ and $\log{S^{\prime}_{ref}} = (1.0, 1.0, 0.0, 0.0, 0.0, 0.0)$.}
\label{fig: sphobfid}
\end{figure*}

Fig.\,\ref{fig: sppbal} shows the asymptotic limit of the entropy second derivative as a function of $\Omega_M$ for given values of the EoS parameters $(w_0, w_a)$. These results might have been anticipated following the argument in \cite{radicella12} where the authors have shown that models with an eternal acceleration comply the laws of thermodynamics. Indeed, we find that $\tilde{S}^{\prime \prime}(a_{max})$ is negative for the fiducial BA model since the DE EoS stays approximately constant thanks to the close to zero $w_a$ value so that the universe tends to a de Sitter\,-\,like state. Moreover, different from the CPL model, the BA EoS has no pathological behaviour in the far future where, on the contrary, the EoS reduces to constant $w_0$. Since we have found $w_0 \sim -1$, we get a $\Lambda$CDM model in the future thus leading to a constant entropy hence $\tilde{S}^{\prime \prime} \sim 0$ which is what Fig.\,\ref{fig: sppbal} shows considering the values of $\tilde{S}_{ref}^{\prime \prime}$ listed in the caption.

\subsection{Hobbit model}

Let us now consider the Hobbit model taking as fiducial model parameters the best fit ones. We obtain the $\tilde{S}^{\prime}_{tot}(a)$ curves in Fig.\,\ref{fig: sphobfid} for different $\tau_{DE}$ values. A comparison with the same plot for the fiducial BA model shows a striking difference with the Hobbit proposal, \textcolor{red}{this model} being able to fulfill the constrain (\ref{eq: spcons}) over the full evolutionary history as long as $\log{\tau_{DE}} > -3.0$ eventually approaching a null value in the asymptotic future. Such a nice behaviour can be qualitatively explained looking at the terms entering Eq.(\ref{eq: sptot}). The only negative term is ${\cal{W}}_1(a) w_{DE}(a)$, while ${\cal{W}}_2(a) w^{2}_{DE}(a) + {\cal{W}}_0$ is positive. When $w_{DE}(a)$ is approximately constant and negative as for the BA model, the balance between the negative and positive terms is regulated by the evolution of the energy densities of the different components. On the contrary, for the Hobbit model, both $w_{DE}(a)$ and the coefficient functions ${\cal{W}}_i(a)$ change in such a way that the net result is a positive derivative. A negative overall value can only be achieved when this balance is not possible anymore which is the case when $\tau_{DE}$ is so small to make the function ${\cal{W}}_1(a)$ much larger than the other factors. Moving to the future, the universe becomes dominated by the Hobbit fluid whose EoS is now approaching the asymptotic limit $w_{DE} = -1$. The universe is therefore driven by a constant $\Lambda$ term which does not evolve anymore thus giving a constant total entropy so that $\tilde{S}^{\prime}_{tot}(a)$ vanishes.

\begin{figure*}
\centering
\includegraphics[width=5.0cm]{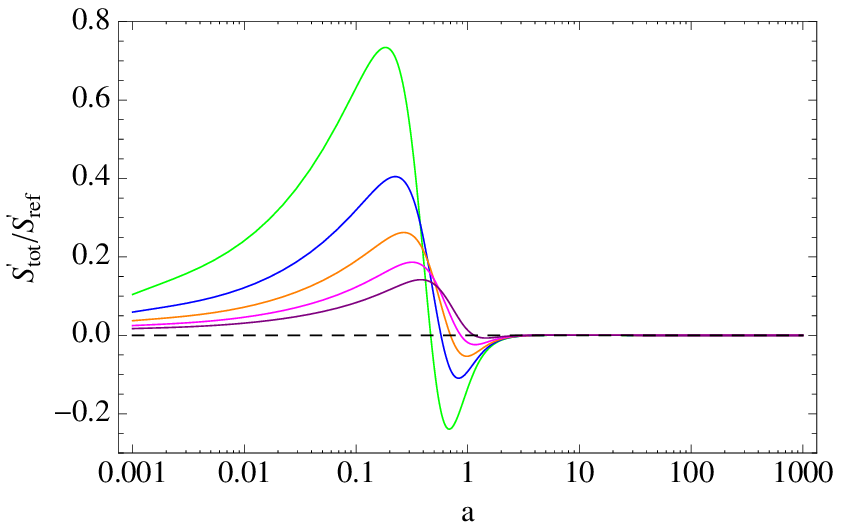}
\includegraphics[width=5.0cm]{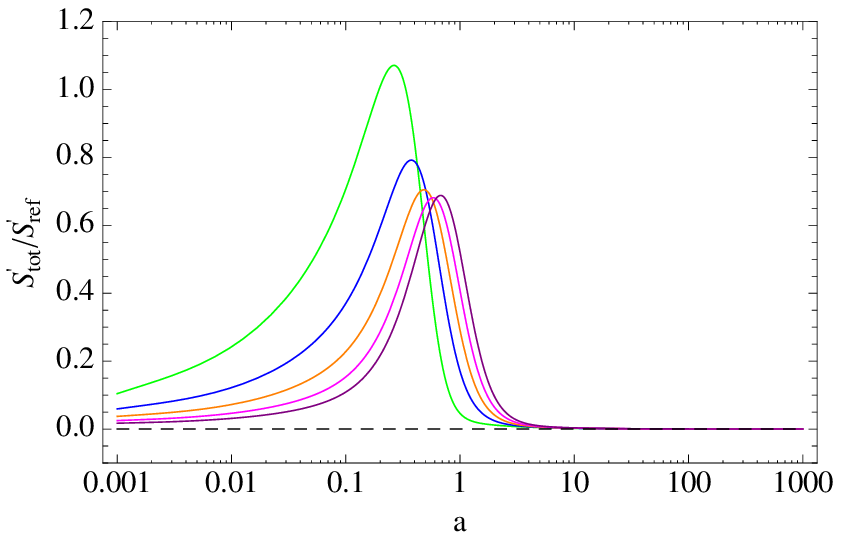}
\includegraphics[width=5.0cm]{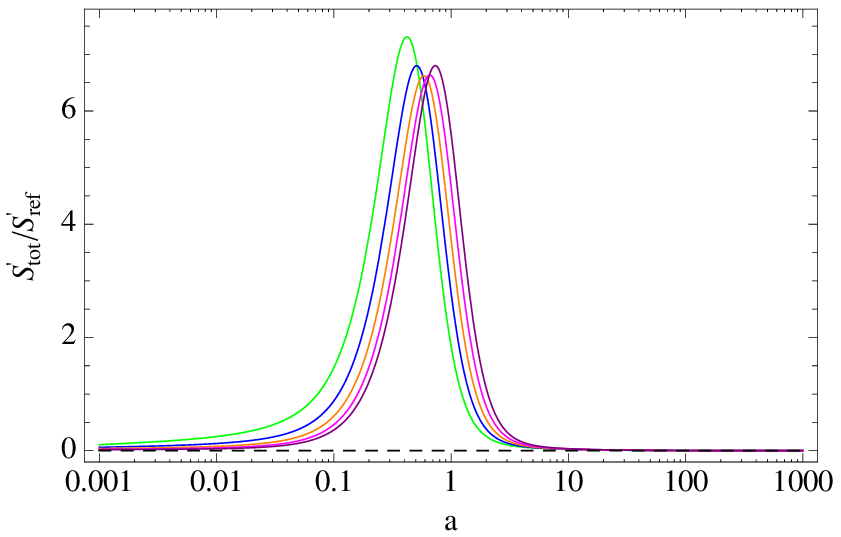} \\
\includegraphics[width=5.0cm]{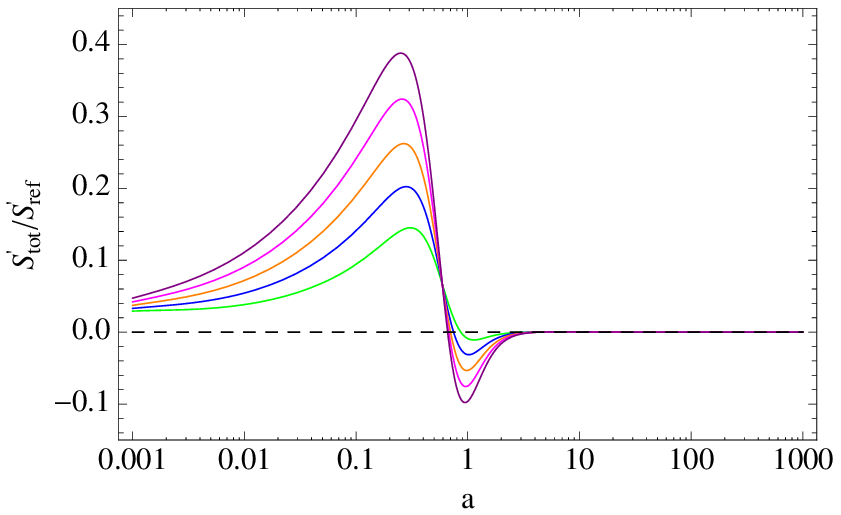}
\includegraphics[width=5.0cm]{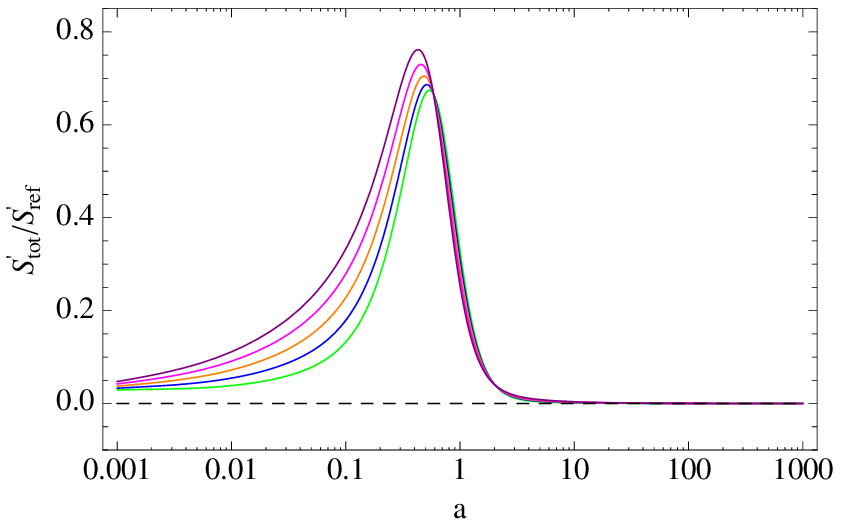}
\includegraphics[width=5.0cm]{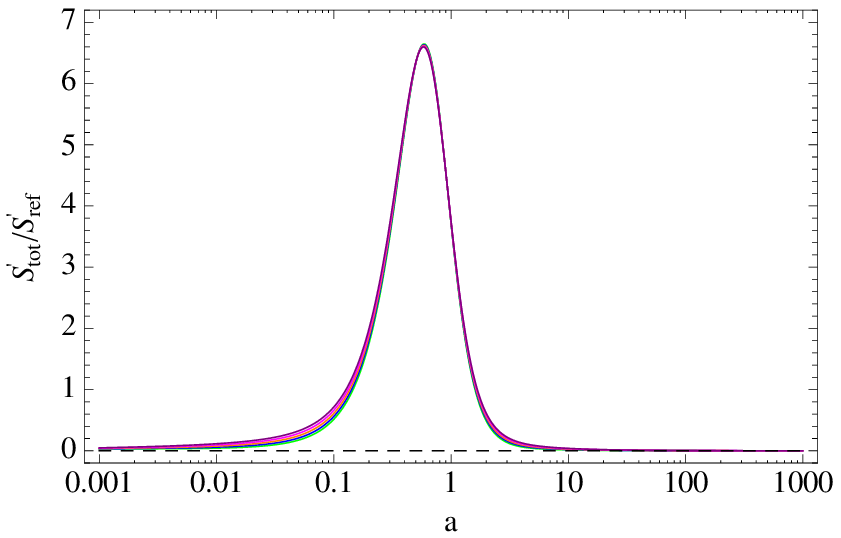} \\
\includegraphics[width=5.0cm]{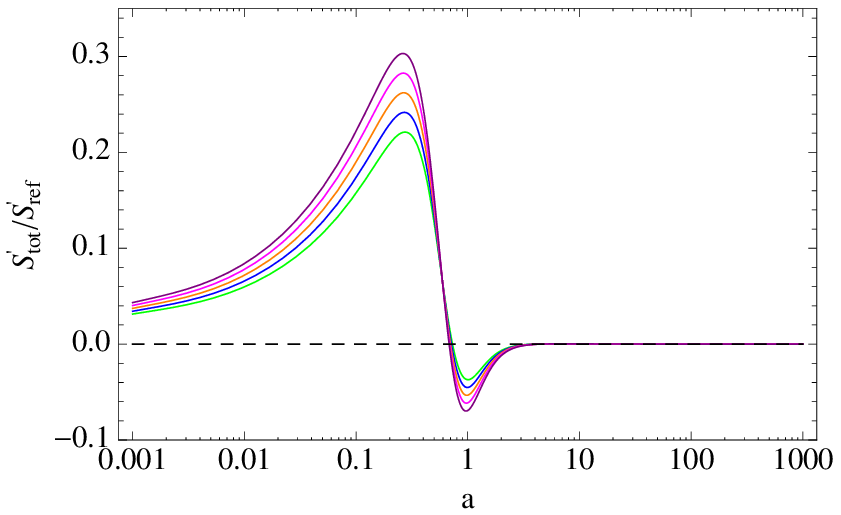}
\includegraphics[width=5.0cm]{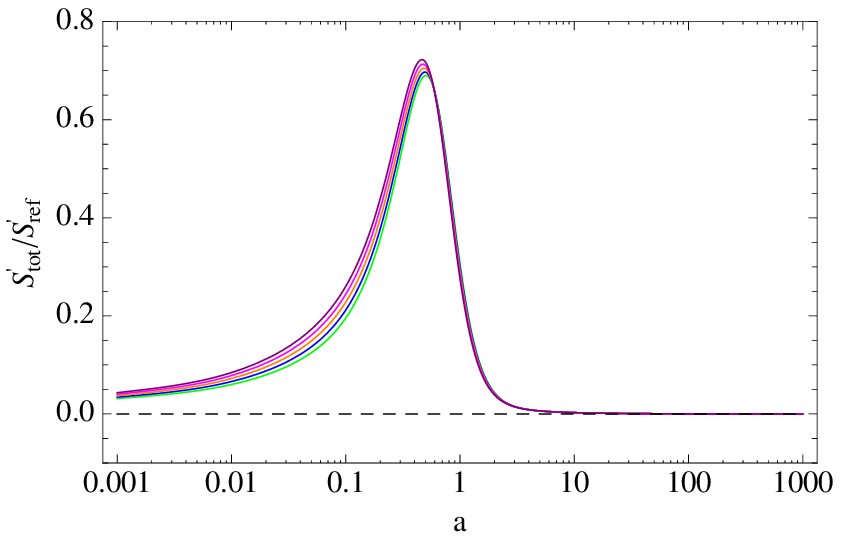}
\includegraphics[width=5.0cm]{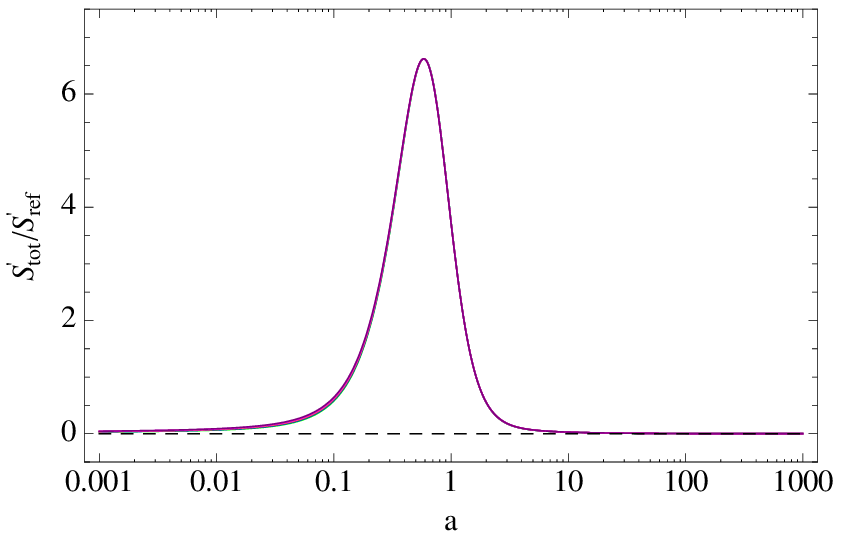} \\
\caption{Scaled first derivative of the entropy for Hobbit models with $\log{\tau_{DE}} = (-3.0, -2.0, -1.0)$ from left to right. Green, blue, orange, magenta, purple lines refers to models with $p/p_{fid} = (0.50, 0.75, 1.00, 1.25, 1.50)$ with $p = (\Omega_M, \alpha, b)$ from top to bottom rows and other parameters set to their fiducial values. Reference values are $\log{S^{\prime}_{ref}} = (2.0, 1.0, 0.0)$ from left to right.}
\label{fig: sphobpar}
\end{figure*}

This qualitative picture is also consistent with what we see in Fig.\,\ref{fig: sphobpar} where we let the model parameter change one at time leaving the others fixed to their fiducial values and exploring three different $\tau_{DE}$ values. First, we note that, no matter the model parameter values, $\tilde{S}^{\prime}_{tot}(a)$ stays always positive provided $\log{\tau_{DE}} > -3.0$ so that we can safely conclude that the Hobbit model fulfills the first thermodynamic constraint provided the present day DE temperature is not too smaller than the horizon one $T_H = H_0 \hbar/k_B$. This is a remarkable result since it suggests that the Hobbit model could be a better physically motivated prescription for the DE evolution with respect to the phenomenological BA model.

Second, we stress that the dependence of $\tilde{S}^{\prime}_{tot}(a)$ on the model parameters is consistent with the qualitative scenario described above. Indeed, models with smaller $\Omega_M$ gives larger $\tilde{S}^{\prime}_{tot}(a)$ values since the smaller is $\Omega_M$, the larger is the Hobbit fluid contribution both in the past and in the present and near future thus explaining why we find larger (in absolute value) peaks. On the other hand, $(\alpha, b)$ control the width of the transition from positive to negative EoS with smaller values leading to models with a smoother $w_{DE}$ and hence a smaller $\tilde{S}^{\prime}_{tot}(a)$. On the contrary, the $(\beta, s)$ parameters have only a marginal impact since they determine the shape of the EoS in the very far past $(a << 1)$ so that their impact is hardly appreciable. 

\begin{figure*}
\centering
\includegraphics[width=5.0cm]{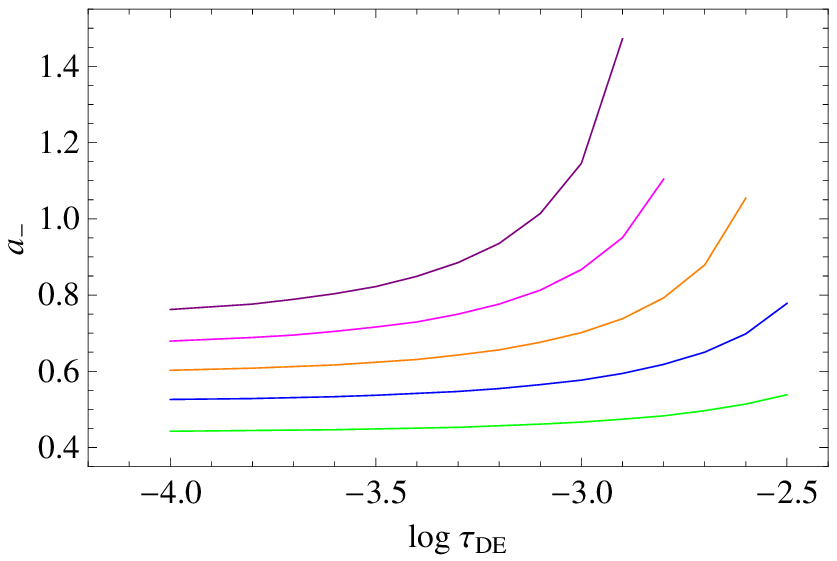}
\includegraphics[width=5.0cm]{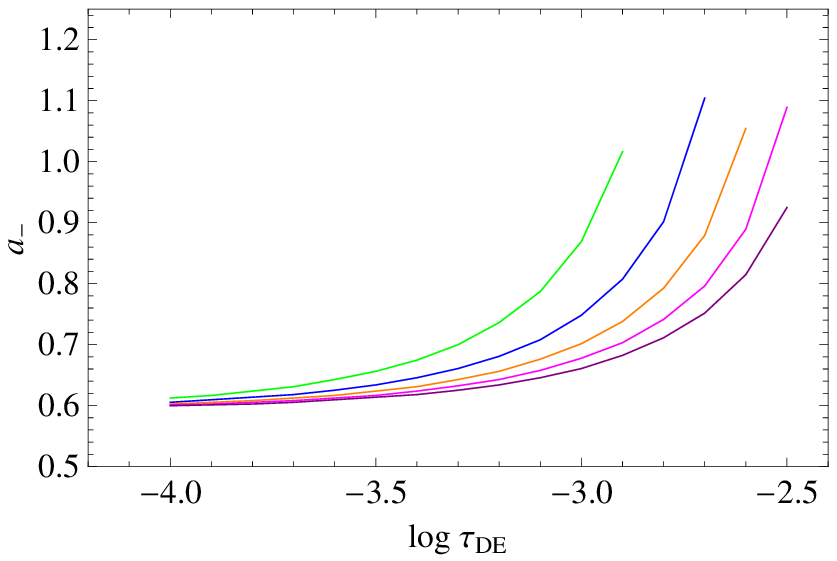}
\includegraphics[width=5.0cm]{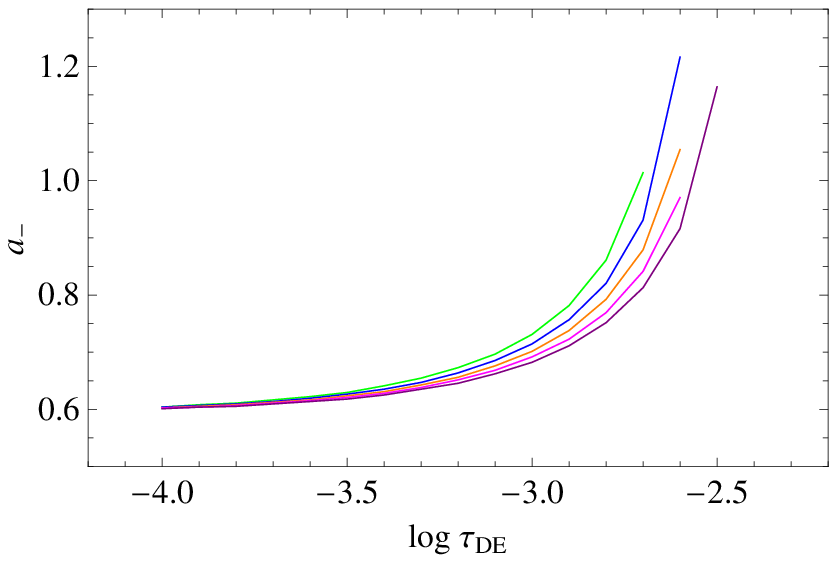} \\
\includegraphics[width=5.0cm]{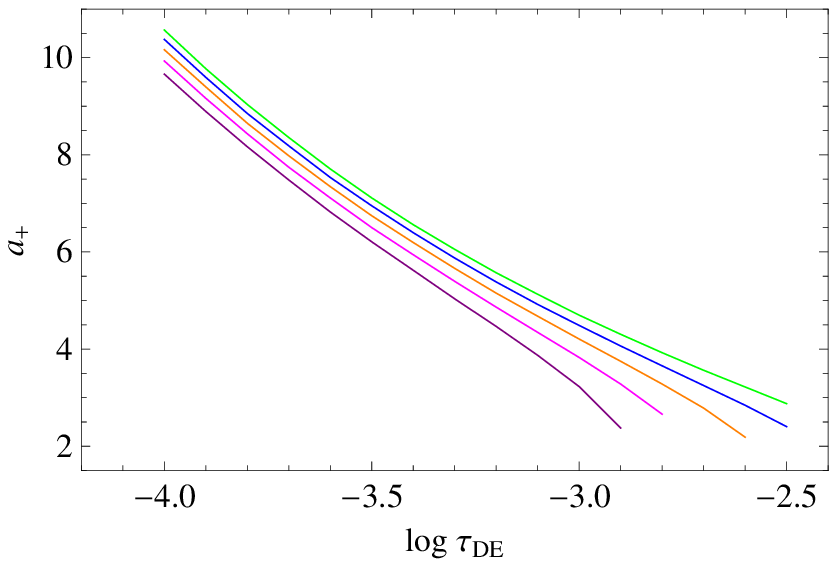}
\includegraphics[width=5.0cm]{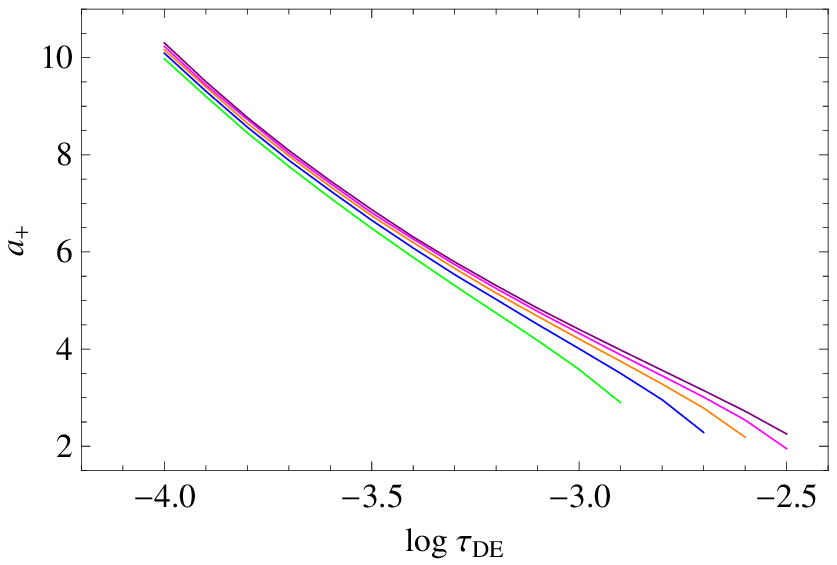}
\includegraphics[width=5.0cm]{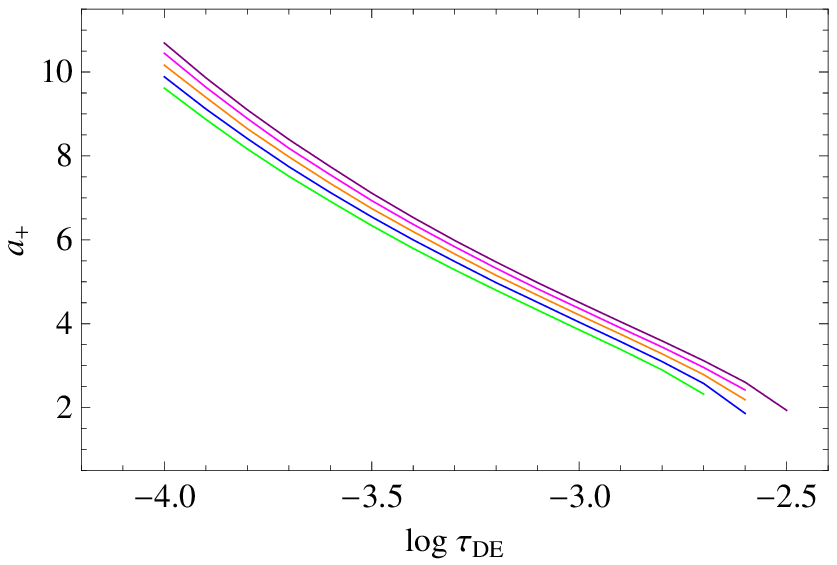} \\
\caption{First (top) and second (bottom) zeros of $\tilde{S}^{\prime}_{tot}(a)$ as function of the DE temperature parameter $\tau_{DE}$ for different model parameters. We fix all parameters to their fiducial values but one with green, blue, orange, magenta, purple lines referring to $p/p_{fid} = (0.50, 0.75, 1.00, 1.25, 1.50)$ and $p = (\Omega_M, \alpha, b)$ from left to right. Note that the curves have different length since there are no zeros for values larger than a critical $\tau_{DE}$.}
\label{fig: zerohob}
\end{figure*}

It is instructive wondering when the entropy derivative violates the constraint (\ref{eq: spcons}). Put in other words, we want to determine the zeros $(a_{-}, a_{+})$ of $\tilde{S}^{\prime}_{tot}(a)$ since it is $\tilde{S}^{\prime}_{tot}(a) \le 0$ for $a_{-} \le a \le a_{+}$. These are shown in Fig.\,\ref{fig: zerohob} as function of $\log{\tau_{DE}}$ for different model parameters combinations. Note that the curves abruptly stop since for $\tau_{DE} > \tau_{crit}$, the function $\tilde{S}^{\prime}_{tot}(a)$ have no zeros being always non negative and asymptoting to zero only for $a \rightarrow \infty$. As can be read from Fig.\,\ref{fig: zerohob}, the critical value depends on the model parameters $(\Omega_M, \alpha, b)$, while we have checked that $(\beta, s, h)$ have no impact at all. In particular, we note that the larger is $\Omega_M$, the smaller is $\tau_{crit}$ in accordance with the qualitative scenario described above. We also note that the larger is $\Omega_M$, the narrower is the range $(a_{-}, a_{+})$, i.e. the smaller is $\Delta a_{\pm} = a_{+} - a_{-}$, so that the better is behaviour of the model from a thermodynamical point of view. Similar considerations argue in favour of models with smaller $(\alpha, b)$ values although they have a minor impact on both $\tau_{crit}$ and $\Delta a_{\pm}$.

\begin{figure*}
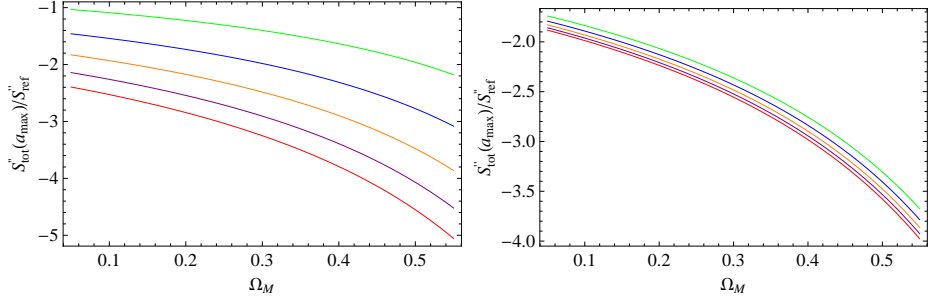

\centering
\includegraphics[width=6.0cm]{spptotvsAlpha.eps}
\includegraphics[width=6.0cm]{spptotvsB.eps} \\
\caption{Asymptotic second derivative of the entropy scaled with respect to a reference value as a function of $\Omega_M$. We set $\tau_{DE} = 1.0$ and all the parameters to their fiducial values but one with with green, blue, orange, magenta, purple lines referring to $p/p_{fid} = (0.50, 0.75, 1.00, 1.25, 1.50)$ and $p = (\alpha, b)$ for left and right panels, respectively. We also set $\log{S^{\prime \prime}_{ref}} = -7.0$ in both panels.}
\label{fig: spphob}
\end{figure*}

Finally, let us look at the asymptotic value of the second order derivative of the total entropy to see whether the Hobbit model fulfills the constraint (\ref{eq: sppcons}). Fig.\,\ref{fig: spphob} shows $\tilde{S}^{\prime \prime}_{tot}(a_{max})$ as a reliable (numerically stable) approximation of this quantity arbitrarily setting $\tau_{DE} = 1$ (but same results are obtained for different values). No matter which are the model parameters values, it turns out that $\tilde{S}^{\prime \prime}_{tot}(a_{max})$ is always negative and quite close to the null value being of order $10^{-7}$. This is consistent with the picture of a universe which is dominated in the future by a DE term with $w_{DE} \simeq -1$, i.e. the Hobbit model reduces to the $\Lambda$CDM one in the asymptotic future. In such a regime, the only term contributing to the entropy is the $\Lambda$\,-\,like one which does not evolve at all. The entropy is therefore constant as a consequence of the universe ending in a thermodynamical equilibrium. Model parameters $(\Omega_M, \alpha, b)$ control how fast this asymptotic state is achieved.

\section{Conclusions}

Recent years have witnessed an impressive progress in observational cosmology with the accumulation of wide and high quality datasets probing both the kinematics and the dynamics of the universe. Such a remarkable progress has nevertheless not been rewarded by a similar advance in the understanding of what is driving the cosmic speed up. Given this frustrating situation, it is worth wondering whether the elusive nature of dark energy can be investigated from a different point of view. In an attempt to add further arrows to hit this rewarding target, we have here tackled dark energy adopting a thermodynamical point of view. We have considered the universe as a closed system and, therefore, we have developed a thermodynamical analysis in analogy with black holes thermodynamics.  As a matter of facts, each cosmological model must not only match the observed data, but also fulfill the generalized second law (GSL) of thermodynamics and the approach to equilibrium. As a consequence, the overall universe entropy must never decrease and reach a maximum in the long run so that the universe asymptotically attains a state of thermodynamical equilibrium. Mathematically, it is the same as asking that the constraints given by Eqs.(\ref{eq: spcons}) and (\ref{eq: sppcons}) are fulfilled. We have therefore first evaluated first and second entropy derivatives with respect to the scale factor $a$ for two different DE models, namely the phenomenological BA parameterization (similar to the popular CPL one) and the empirically motivated Hobbit scenario. In practice,  after having shown that both are in agreement with present day data (which also allows us to constrain their parameters), we have then checked whether they pass the GSL constraints with green light. 

It turns out that this is actually possible for both models, but a critical role is played by the unknown value of the present day DE temperature. As a general result, we find that, for the BA model, the first derivative of the total entropy $\tilde{S}^{\prime}_{tot}(a)$ changes sign at least one time, i.e. the equation $\tilde{S}^{\prime}_{tot}(a) = 0$ has at least one solution. Depending on $\tau_{DE}$ and the model parameters $(\Omega_M, w_0, w_a)$ values, $\tilde{S}^{\prime}_{tot}(a)$ can then become positive again or stays negative converging to a close to null limit in the far future. It is worth noting that as far as $\log{\tau_{DE}} > 0$, $\tilde{S}^{\prime}_{tot}(a)$ stays positive over the full evolutionary history of the universe provided model parameters are suitably chosen yet remaining within the $68\%$ CL allowed by the data. On the other hand, the Hobbit model can be reconciled with the GSL quite easily as far as the condition $\log{\tau_{DE}} > -2.5$ is imposed on the present day DE temperature. We indeed find that the function $\tilde{S}^{\prime}_{tot}(a)$ has no zeros in this case, while the second derivative of the entropy always asymptotes to an almost null value no matter which values are set for $\tau_{DE}$ and the model parameters $(\Omega_M, \alpha, \beta, b, s)$. This nice thermodynamical behaviour originates from the characteristics of the Hobbit EoS which makes the fluid mimic the cosmological  constant $\Lambda$ in the far future (so that the entropy stays constant and the universe is in a state of thermodynamical equilibrium).

It is somewhat ironic that the thermodynamical analysis points at the Hobbit model as a better physically motivated scenario, but it is the BA EoS which comes out as the preferred one from the fitting procedure because of the lower number of parameters (being the likelihood almost the same for the two models). Such a contradictory result is actually an outcome of how the two models have been conceived. The BA parameterization aims at generalizing the $\Lambda$CDM model and it has been designed so that it can efficiently mimic a wide class of DE models including the (unknown) one underlying the data themselves. It is therefore not surprising that it is able to fit the data especially considering that the best fit parameters are indeed quite close to those of the concordance $\Lambda$CDM scenario. On the contrary, the Hobbit model has been manufactured by hand so that it mimic standard fluids in the past and the cosmological constant in the recent and future epochs. Since each one of these fluids correctly behaves from a thermodynamical point of view, we expect that the Hobbit one does the same except in the transition regime from matter to the $\Lambda$ term. 

Although the above argument explains why the two independent indicators point in different directions, it is clearly desirable to have both tests select the same model. To this end, one could importance sample the MCMC chains of a given model (not necessarily those we have investigated here) so that the final selected parameters identify configurations that both match the data and fulfill the thermodynamic constraints. Two difficulties are in order here. First, both Eqs.(\ref{eq: spcons}) and (\ref{eq: sppcons}) are inequalities so that they can not be implemented as a likelihood function to be maximized. On the contrary, for each set of parameters along the chain, one must compute the entropy derivatives and check their sign all over the range $(a_{min}, a_{max})$. Only the configurations passing the GSL constraints are then saved thus significantly reducing the parameter space narrowing down the confidence ranges. However, such an approach is straightforward if all the quantities entering Eqs.(\ref{eq: sptot}) and (\ref{eq: spptot}) are analytical otherwise one should compute $\tilde{S}^{\prime}_{tot}(a)$ and 
$\tilde{S}^{\prime \prime}_{tot}(a)$ numerically which can be quite time consuming (especially if one needs a long chain to achieve MCMC convergence). Actually, this is only a numerical problem which can likely be solved by a suitable algorithm or running the code on fast machines. The most serious hurdle is indeed represented by the ignorance on the DE temperature, i.e., the value of the $\tau_{DE}$ parameter. As we have shown for both the BA and Hobbit models, for a given set of model parameters, the number of zeros of the function $\tilde{S}^{\prime}_{tot}(a)$ depend on $\tau_{DE}$. Put in other words, each set along the MCMC chain can be accepted or rejected according to $\tau_{DE}$ being smaller or larger than a critical value. As a mandatory step before implementing a joint data\,-\,GSL analysis, one must find a way to constrain the present day DE temperature.

Although some significant work is needed for a more efficient implementation, the results in this preliminary analysis show that thermodynamics can offer an additional yet competitive way to compare different DE models. The conjecture that universe evolutions can be depicted from the thermodynamical point of view in analogy with black holes allows to settle two natural constraints that have to be satisfied along the cosmological history.  In our opinion, adding more and more data is a rewarding strategy to constrain model parameters but adopting suitable conceptual paradigms can allow to further improve the differentiation process among rival competing approaches.

\acknowledgments

VFC is funded by Italian Space Agency (ASI) through contract Euclid\,-\,IC (I/031/10/0) and acknowledge financial contribution from the agreement ASI/INAF/I/023/12/0. NR is supported by INFN ({\sl Iniziativa Specifica} QGSKY) and acknowledges the COST Action CA15117 (CANTATA).

\end{document}